\documentclass[aps,prb,twocolumn,superscriptaddress,amsmath,amssymb]{revtex4}
\usepackage{graphicx}
\usepackage{amsmath}
\usepackage{float}
\usepackage{color}
\usepackage{appendix}
\usepackage[normalem]{ulem}
\usepackage[colorlinks,citecolor=blue,urlcolor=blue]{hyperref}

\begin{document}

\title{Non-Bloch topological invariants in a non-Hermitian domain-wall system}

\author{Tian-Shu Deng}
\affiliation{CAS Key Laboratory of Quantum Information, University of Science and Technology of China, Hefei, Anhui, 230026, China}
\affiliation{CAS Center For Excellence in Quantum Information and Quantum Physics}
\author{Wei Yi}
\email{wyiz@ustc.edu.cn}
\affiliation{CAS Key Laboratory of Quantum Information, University of Science and Technology of China, Hefei, Anhui, 230026, China}
\affiliation{CAS Center For Excellence in Quantum Information and Quantum Physics}

\date{\today}
\begin{abstract}
We study non-Bloch bulk-boundary correspondence in a non-Hermitian Su-Schieffer-Heeger model in a domain-wall configuration where the left and right bulks have different parameters. Focusing on the case where chiral symmetry is still conserved, we show that non-Hermitian skin effects of bulk states persist in the system, while the definition of the non-Bloch winding number of either bulk depends on parameters on both sides of the boundary. Under these redefined non-Bloch topological invariants, we confirm non-Bloch bulk-boundary correspondence under the domain-wall configuration, which exemplifies the impact of boundary conditions in non-Hermitian topological systems.
\end{abstract}
\maketitle

\section{Introduction}

A prominent feature of topological matter is the existence of robust topological edge states at boundaries, whose existence are related to bulk topological invariants according to the principle of bulk-boundary correspondence~\cite{topo1,topo2,topo3,topo4,topo5,topo6,topo7}. In recent studies, it has been shown that bulk-boundary correspondence in some non-Hermitian topological systems should be modified in order to correctly predict the number of topological edge states at their boundaries~\cite{NHtopo1,NHtopo2,NHtopo3,NHtopo4,NHtopo5,NHtopo6,NHtopo8,NHtopo10,NHtopo11,NHtopo115,NHtopo12,NHtopo13,NHtopo14,NHtopo17,NHtopo18}. Such a phenomenon is closely related to the non-Hermitian skin effect, where bulk states in the corresponding non-Hermitian systems become localized at boundaries~\cite{NHtopo3,NHtopo4,NHtopo8,NHtopo10,NHtopo13}. The deviation of bulk-state wave functions from extended Bloch waves subsequently necessitates the extension of the Brillouin zone onto the complex plane, where non-Bloch topological invariants capable of accounting for topological edge states can be defined~\cite{NHtopo3,NHtopo4}.

Non-Hermitian skin effects and non-Bloch topological invariants significantly extend the conventional understanding of bulk-boundary correspondence, and are of fundamental theoretical importance for a deeper understanding of topological phenomena in non-Hermitian systems. In light of recent advances in engineering non-Hermitian topology in synthetic systems~\cite{NHsyn6,NHsyn1,NHsyn2,NHsyn5,NHsyn3,NHsyn4,NHsyn7,NHsyn8}, the study of bulk-boundary correspondence in non-Hermitian settings is also experimentally relevant. In previous studies of non-Bloch topological invariants however, the focus has been on non-Hermitian topological systems with open boundary conditions~\cite{NHtopo3,NHtopo4,NHtopo12}. Experimentally, a common alternative is a domain-wall configuration, where two bulks with distinct parameters are in contact through a common boundary. To construct non-Bloch topological invariants in this case requires a complete understanding of bulk-state wave functions, which could be quite different from those under the open boundary condition.

In this work, we study non-Bloch bulk-boundary correspondence in a non-Hermitian Su-Schieffer-Heeger (SSH) model\cite{NHtopo1,NHtopo3,NHtopo5,GeoSSH,NHSSH} under a domain-wall configuration where the left and right bulks have different parameters. We focus on a model where non-Hermitian skin effects are present and the conventional Bloch bulk-boundary correspondence breaks down. Importantly, we show that localized bulk wave functions on either side of the boundary are affected by parameters of both bulks. Because of this complication, for the left or the right bulk alone, one can construct two different generalized Brillouin zones based on localized bulk wave functions. We prove that different generalized Brillouin zones for a given bulk yield the same non-Bloch topological invariant, which, together with the non-Bloch topological invariant of the other bulk, yields the correct bulk-boundary correspondence of the domain-wall configuration. We provide a
detailed characterization of the evolution of the energy spectrum, winding numbers and bulk- and edge-state wave functions with varying system parameters. In particular, we systematically study the effects of the size of bulks on non-Bloch winding numbers.
Our work explicitly reveals the impact of the domain-wall boundary condition on the non-Bloch bulk-boundary correspondence, and provides a solid foundation for future experimental studies of non-Bloch topological invariants.

The paper is organized as follows. In Sec. II, we present the model Hamiltonian as well as the Bloch topological invariants. We then study bulk-state wave functions in Sec. III, which allow us to construct the generalized Brillouin zones and non-Bloch winding numbers for each bulk. In Sec. IV, we confirm the the non-Bloch bulk-boundary correspondence by numerically characterizing the energy spectrum, and bulk- and edge-state wave functions. We then study in detail the impact of the length ratio of two bulks on the energy spectrum and the non-Bloch winding numbers in Sec. V. We summarize in Sec. VI.

\section{Model and Bloch topological invariants}
\begin{figure}[tbp]
\center{\includegraphics[width=0.8\linewidth]{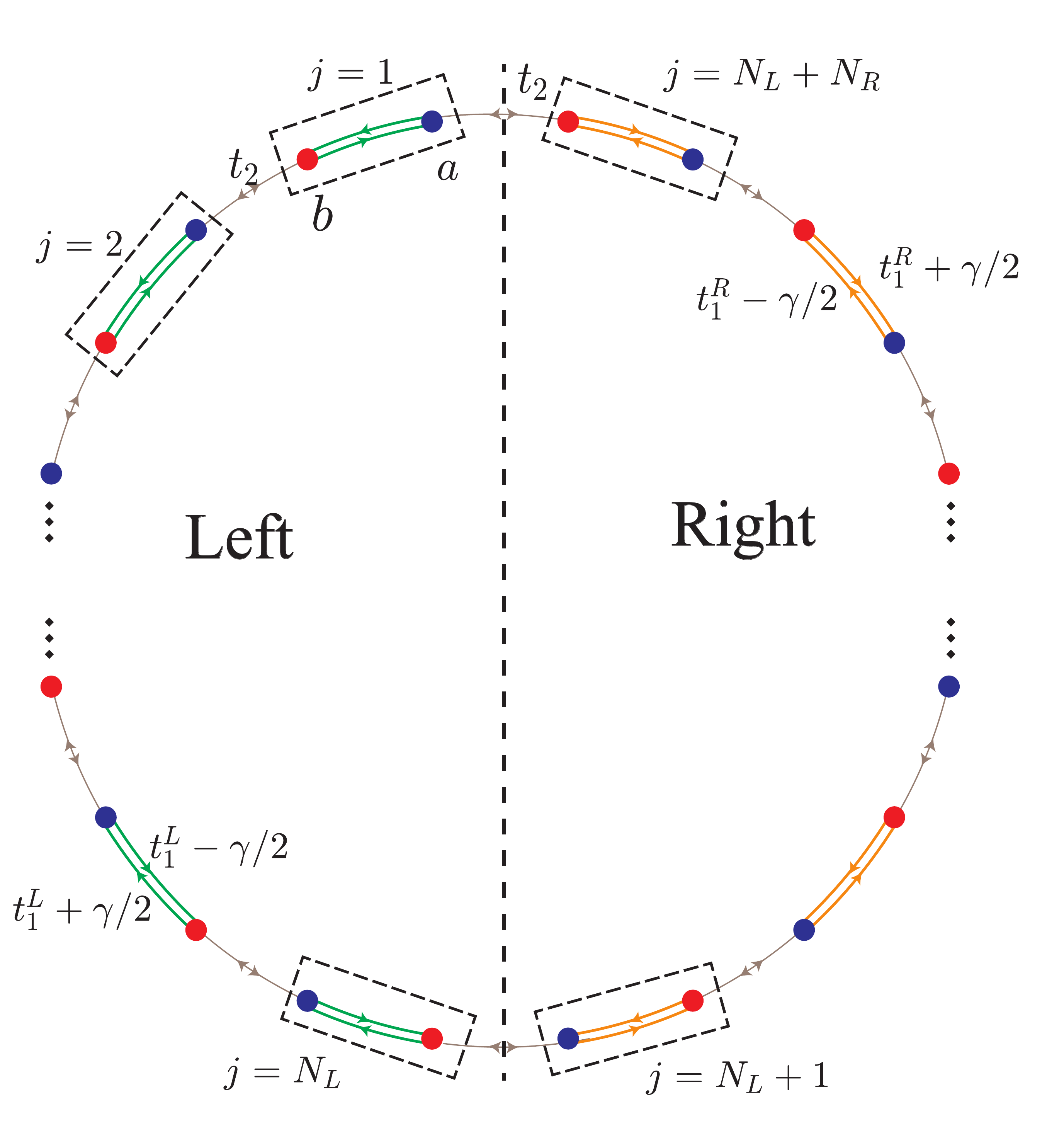}}
\caption{Illustration of a non-Hermitian SSH model under the domain-wall configuration. The intra-cell hopping from sublattice sites $a$ to $b$ in the left(right) bulk is $t^{L(R)}_1-\gamma/2$, while the intra-cell hopping from $b$ to $a$ is $t^{L(R)}_1+\gamma/2$. For simplicity, we assume both bulks have the same $t_2$ and $\gamma$.}
\label{fig:fig1}
\end{figure}

As illustrated in Fig.~\ref{fig:fig1}, we consider a non-Hermitian SSH model in a domain-wall configuration on a ring. The Hamiltonian can be written as
\begin{equation}
\label{HLR}
H=H_{L}+H_{R},
\end{equation}
where
\begin{align}
\label{Halpha}
H_{\alpha}&=\sum_{j\in J_{\alpha}}(t_{1}^{\alpha}+\frac{\gamma}{2})a_{j}^{\dagger}b_{j}+(t_{1}^{\alpha}-\frac{\gamma}{2})b_{j}^{\dagger}a_{j}\nonumber\\
&+t_{2}a_{j+1}^{\dagger}b_{j}+t_{2}b_{j}^{\dagger}a_{j+1}.
\end{align}
Here, $\alpha=(L,R)$ denotes the left or right bulk, $a_j^\dagger$ ($a_j$) and $b_j^\dagger$ ($b_j$) are respectively the creation (annihilation) operators for the sublattice sites $a$ and $b$ on the $j$-th unit cell. The left (right) bulk contains  $N_L$ ($N_R$) unit cells, which we label as $J_{L}=\{x\in \mathbb{Z}|1\leqslant x\leqslant N_{L}\}$ and $J_{R}=\{x\in \mathbb{Z}|N_{L+1}\leqslant x\leqslant N_{R}+N_{L}\}$, respectively. Since the system has a ring geometry, we take $a_{N_{L}+N_{R}+1}^{\dagger}(a_{N_{L}+N_{R}+1})=a_{1}^{\dagger}(a_{1})$ and $b_{N_{L}+N_{R}+1}^{\dagger}(b_{N_{L}+N_{R}+1})=b_{1}^{\dagger}(b_{1})$. The intra-cell hopping difference $\gamma$ controls the non-Hermicity of the system, such that when $\gamma=0$ the Hamiltonian is reduced to a conventional Hermitian SSH model.
The non-Hermitian SSH model has chiral symmetry $\Gamma H\Gamma^{-1}=-H$, with the chiral-symmetry operator $\Gamma=\sum_{j=1}^{N_{L}+N_{R}}(a_{j}^{\dagger}a_{j}-b_{j}^{\dagger}b_{j})$.

Following the conventional definition of winding numbers in the Hermitian case, one can define the Bloch winding numbers from the Bloch Hamiltonians $h_{\alpha}(k)$ of the two bulks, where
\begin{equation}
\label{Hksigma}
h_{\alpha}(k)=h^{x}_{\alpha}(k)\sigma_{x}+h^{y}_{\alpha}(k)\sigma_{y}.
\end{equation}
Here, $h^{x}_{\alpha}=t^\alpha_{1}+t_{2}\cos k$ and $h^{y}_{\alpha}=t_{2}\sin k+\frac{\gamma}{2}i$, and
$\sigma_x$ and $\sigma_y$ are Pauli matrices. The corresponding eigen-energy spectrum is
\begin{equation}
\label{Ek}
E_{\alpha}(k)=\pm\sqrt{(t_{1}^{\alpha}+t_{2}\cos k)^{2}+(t_{2}\sin k+\frac{\gamma}{2}i)^{2}},
\end{equation}
where the energy gap closes at the exceptional points $t_{1}^{\alpha}=\pm t_{2}\pm\frac{\gamma}{2}$.

The Bloch winding numbers for the two bulks are then given by
\begin{equation}
\label{nu}
\nu_{\alpha}=\frac{1}{\pi}\int_{0}^{2\pi}\frac{\langle\chi_{\alpha}|i\partial_{k}|\psi_{\alpha}\rangle}{\langle\chi_{\alpha}|\psi_{\alpha}\rangle}dk,
\end{equation}
where $|\psi_\alpha\rangle$ ($|\chi_\alpha\rangle$) is the right (left) eigenvector of $h_{\alpha}(k)$, with
$h_{\alpha}(k)|\psi_{\alpha}\rangle=E_\alpha(k)|\psi_{\alpha}\rangle$ and
$h_{\alpha}^\dagger(k)|\chi_{\alpha}\rangle=E_\alpha^{*}(k)|\chi_{\alpha}\rangle$.
However, due to the non-Hermitian skin effects of bulk states, one cannot predict the correct number of zero modes in the domain-wall configuration from $\nu_{R}-\nu_{L}$. For that purpose, we need to use the non-Bloch winding numbers, which require an understanding of bulk-state wave functions.

Finally, substituting the explicit forms of $|\psi_\alpha\rangle$ and $|\chi_\alpha\rangle$ into Eq.~(\ref{nu}), we have
\begin{equation}
\label{nualpha}
\nu_{\alpha}=\frac{1}{2\pi}\int dk\frac{-h_{\alpha}^{x}\frac{\partial h_{\alpha}^{y}}{\partial k}+h_{y}^{\alpha}\frac{\partial h_{\alpha}^{x}}{\partial k}}{(h_{\alpha}^{x})^{2}+(h_{\alpha}^{y})^{2}},
\end{equation}
which will be useful later for the construction of non-Bloch winding numbers

\section{Bulk-state wave functions}
We study bulk-state wave functions by writing down the eigenstates as
\begin{equation}
\label{psi}
|\Psi\rangle=\sum_{j=1}^{N_{L}+N_{R}}\left(\psi_{a,j}a_{j}^{\dagger}+\psi_{b,j}b_{j}^{\dagger}\right)|0\rangle,
\end{equation}
where $\psi_{a(b),j}$ are the on-site wave functions. Substituting Eq.~(\ref{HLR}), Eq.~(\ref{Halpha}) and Eq.~(\ref{psi}) into the Schr\"odinger's equation $H|\Psi\rangle=E|\Psi\rangle$, we obtain the recurrence relation
\begin{align}
\label{psialpha1}
E\psi_{a,j}&=(t_{1}^{\alpha}+\frac{\gamma}{2})\psi_{b,j}+t_{2}\psi_{b,j-1},\\
\label{psialpha2}
E\psi_{b,j}&=(t_{1}^{\alpha}-\frac{\gamma}{2})\psi_{a,j}+t_{2}\psi_{a,j+1},
\end{align}
from which we have
\begin{align}\label{psiaabb}
&t_{2}(t_{1}^{\alpha}+\frac{\gamma}{2})\psi_{a(b),j+1}+t_{2}(t_{1}^{\alpha}-\frac{\gamma}{2})\psi_{a(b),j-1}\nonumber\\
&+(t_{1}^{\alpha}{}^{2}-\frac{\gamma^{2}}{4}+t_{2}^{2}-E^{2})\psi_{a(b),j}=0.
\end{align}

Apparently, $\psi_{a,j}$ and $\psi_{b,j}$ are decoupled from each other. It is then possible to write the general solution for bulk states as
\begin{equation}
\label{psiL}
\left(\begin{array}{c}
\psi_{a,j}\\
\psi_{b,j}
\end{array}\right)=\left(\begin{array}{c}
\phi_{a}^{(1)}\\
\phi_{b}^{(1)}
\end{array}\right)\lambda_{L,1}^{j}+\left(\begin{array}{c}
\phi_{a}^{(2)}\\
\phi_{b}^{(2)}
\end{array}\right)\lambda_{L,2}^{j},
\end{equation}
for $j\in J_{L}$ and
\begin{equation}
\label{psiR}
\left(\begin{array}{c}
\psi_{a,j}\\
\psi_{b,j}
\end{array}\right)=\left(\begin{array}{c}
\varphi_{a}^{(1)}\\
\varphi_{b}^{(1)}
\end{array}\right)\lambda_{R,1}^{j-N_{L}}+\left(\begin{array}{c}
\varphi_{a}^{(2)}\\
\varphi_{b}^{(2)}
\end{array}\right)\lambda_{R,2}^{j-N_{L}},
\end{equation}
for $j\in J_{R}$. Here, coefficients $\phi_{a,b}^{(i)}$ ,$\varphi_{a,b}^{(i)}$ satisfy
\begin{align}
\label{fgi1}
&\frac{\phi_{a}^{(i)}}{\phi_{b}^{(i)}}=\frac{E}{(t_{1}^{L}-\frac{\gamma}{2})+t_{2}\lambda_{L,i}}=f_i,\\
\label{fgi2}
&\frac{\varphi_{a}^{(i)}}{\varphi_{b}^{(i)}}=\frac{E}{(t_{1}^{R}-\frac{\gamma}{2})+t_{2}\lambda_{R,i}}=g_i,
\end{align}
while $\lambda_{\alpha,i}$ satisfy the following characteristic equation of the linear recurrence relation Eq.~(\ref{psiaabb})
\begin{equation}
\label{beta}
t_{2}(t_{1}^{\alpha}+\frac{\gamma}{2})\lambda_{\alpha,i}+t_{2}(t_{1}^{\alpha}-\frac{\gamma}{2})\frac{1}{\lambda_{\alpha,i}}+(t_{1}^{\alpha})^{2}-\frac{\gamma^{2}}{4}+t_{2}^{2}-E^{2}=0,
\end{equation}
where $i=(1,2)$ denote different roots of Eq.~(\ref{beta}).

Substituting Eqs.~(\ref{psiL}-\ref{fgi2}) into the Schr\"odinger's equation at the boundary, we derive a set of linear equations for the coefficients $\left[\phi_{b}^{(1)},\phi_{b}^{(2)},\varphi_{b}^{(1)} ,\varphi_{b}^{(2)}\right]$ (see Appendix~\ref{BC}). Sending the determinant of the coefficient matrix to zero, we have
\begin{align}
\label{detM}
&\frac{(1-\lambda_{L,1}^{N_{L}}\lambda_{R,1}^{N_{R}})(1-\lambda_{L,2}^{N_{L}}\lambda_{R,2}^{N_{R}})}{(\lambda_{L,1}f_{1}-\lambda_{R,1}g_{1})(\lambda_{L,2}f_{2}-\lambda_{R,2}g_{2})}\nonumber\\
&=\frac{(1-\lambda_{L,1}^{N_{L}}\lambda_{R,2}^{N_{R}})(1-\lambda_{L,2}^{N_{L}}\lambda_{R,1}^{N_{R}})}{(\lambda_{L,1}f_{1}-\lambda_{R,2}g_{2})(\lambda_{L,2}f_{2}-\lambda_{R,1}g_{1})}.
\end{align}

While $\lambda_{\alpha,i}$ can be regarded as functions of eigen-energy $E$ through Eq.~(\ref{beta}), Eq.~(\ref{detM}) gives all bulk-state eigen-energies. Further, $\lambda_{\alpha,i}$ give the non-Bloch description of the real-space wave functions, which are crucial for constructing the non-Bloch Brillouin zones. For simplicity, we first assume $N_R=N_L=N$ and will discuss the case of $N_R\neq N_L$ in Sec.~\ref{rN}.
For $N_R=N_L=N$ and in the thermodynamic limit with $N\rightarrow \infty$, solutions of $\lambda_{\alpha,i}$ satisfy
\begin{equation}
\label{zetaE}
\zeta(\lambda_{L,1},\lambda_{L,2},\lambda_{R,1},\lambda_{R,2})=0,
\end{equation}
where the specific form and derivation of $\zeta$ is given in Appendix~\ref{zetaf}.

\begin{figure}[tbp]
\center{\includegraphics[width=1\linewidth]{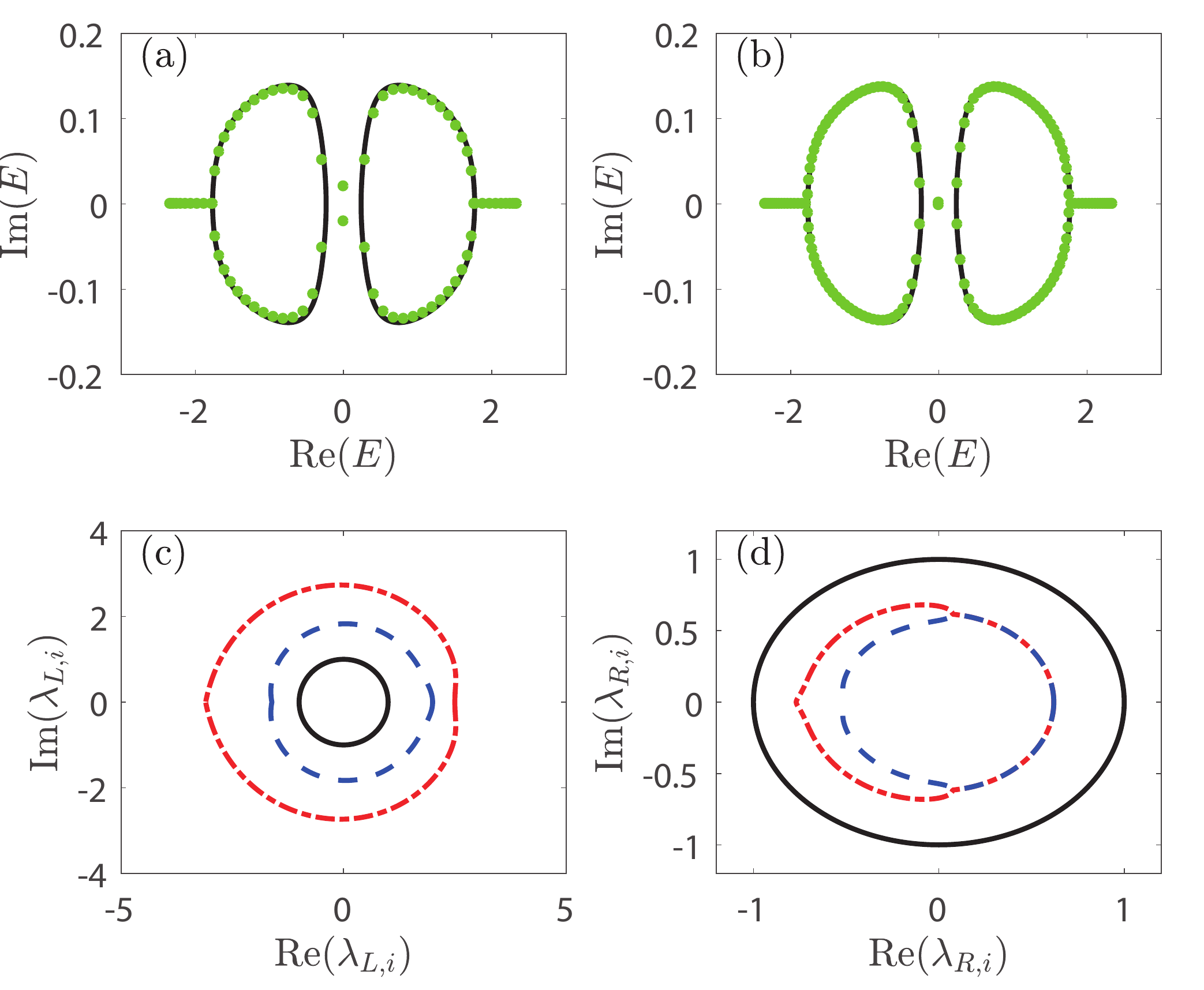}}
\caption{ (a)(b) Theoretical (black lines) and numerical (green dots) results of bulk-state energy spectrum.
The length of chain is $N_L=N_R=20$ for (a) and $N_L=N_R=40$  for (b). (c) Non-Bloch Brillouin zones of the left bulk, represented by $\lambda_{L,1}$ (red dash-dotted line) and $\lambda_{L,2}$(blue dashed line) on the complex plane. (d) Non-Bloch Brillouin zones of the right bulk, represented by $\lambda_{R,1}$ (red dash-dotted line) and $\lambda_{R,2}$(blue dashed line) on the complex plane. In (c)(d), we take the thermodynamic limit $N_L=N_R\rightarrow \infty$, and we also plot the Bloch Brillouin zones with black solid lines for comparison. For all subplots, we take $t_1^L=-t_2$, $t_1^R=1.5t_2$, $\gamma=1.33t_2$.}
\label{fig:fig2}
\end{figure}

Eq.~(\ref{zetaE}) can be regarded as an equation with a single complex variable $E$, as $\lambda_{\alpha,i}$ are all related to $E$ through Eq.~(\ref{beta}). Solution of Eq.~(\ref{zetaE}) thus determines the bulk energy spectrum. This is confirmed in Fig.~\ref{fig:fig2}(a)(b), where the solution of Eq.~(\ref{zetaE}) agrees well with the numerically calculated energy spectrum of the domain-wall system. More importantly, as $E$ varies, the trajectories of $\lambda_{\alpha,i}$ on the complex plane can be related to generalized Brillouin zones.
In Fig.~\ref{fig:fig2}(c)(d), we plot $\lambda_{\alpha,i}$ in the complex plane. In the Hermitian limit ($\gamma=0$), $\lambda_{\alpha,i}$ are all located on a unit circle on the complex plane. Thus one can identify them as $\lambda_{\alpha,1}=e^{ik}$ and $\lambda_{\alpha,2}=e^{-ik}$, in which case Eqs.~(\ref{psiL}) and (\ref{psiR}) are reduced to the Bloch wave functions, and $\arg(\lambda)$ correspond to quasi-momenta $k$ in the Brillouin zone. In contrast, when the Hamiltonian becomes non-Hermitian, $\lambda_{\alpha,i}$ are deformed from the unit circle, which give rise to deformed Brillouin zones. Different from a non-Hermitian SSH model with an open boundary condition, under the domain-wall configuration that we consider here, one can define four different deformed Brillouin zones. Nevertheless, in the next section, we will show that non-Bloch bulk-boundary correspondence can still be established based on winding numbers defined on these deformed Brillouin zones.

\section{Non-Bloch topological  invariants}

\begin{figure*}[tbp]
\center{\includegraphics[width=1\linewidth]{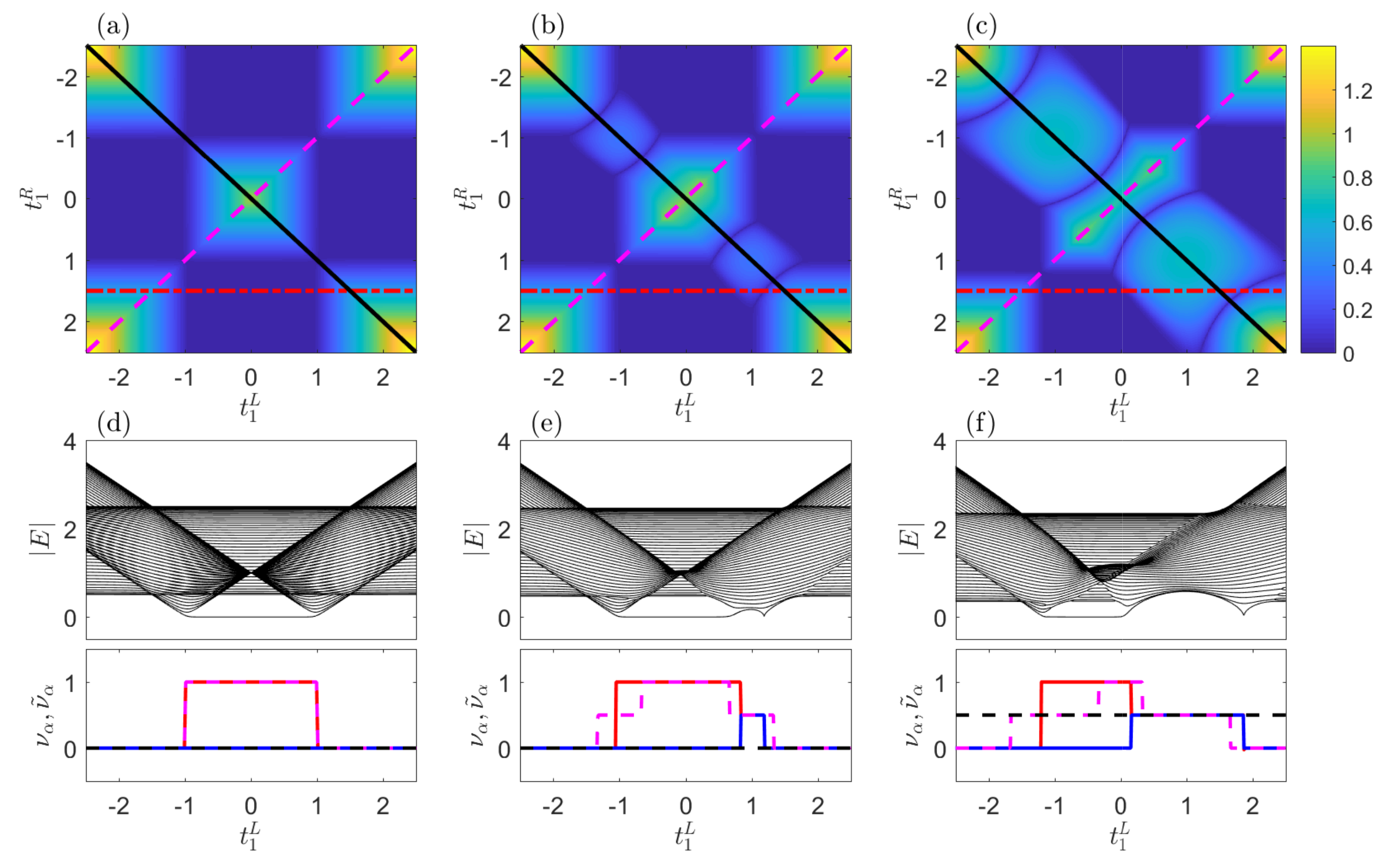}}
\caption{(a)(b)(c) Contour plots of absolute values of the energy-spectrum minimum on the $t_1^L$--$t_1^R$ plane.
Here, we take $N_L=N_R=40$. We also take $\gamma=0$ for (a), $\gamma=0.67t_2$ for (b) and $\gamma=1.33t_2$ for (c). The black solid lines are given by $t_1^L=t_1^R$, where the domain-wall configuration is reduced to single homogeneous bulk with a periodic boundary condition. The magenta dashed lines are given by $t_1^L=-t_1^R$, where the bulk-state wave functions satisfy $|\lambda_{L,1}|=|\lambda_{L,2}|$, the same as those under the open-boundary condition. The red dashed-dotted line in (a)(b)(c) correspond to parameters we use in (d)(e)(f) with $t_1^R=1.5t_2$. (d)(e)(f) The absolute values of the energy spectrum (upper panels) and various winding numbers (lower panel). In the lower panel, we show the Bloch winding numbers for the left bulk $\nu_L$ (magenta dashed lines) and the right bulk $\nu_R$ (black dashed lines), as well as non-Bloch winding numbers for the left bulk $\tilde{\nu}_L$ (red solid lines) and  the right bulk $\tilde{\nu}_R$ (blue solid lines).}
\label{fig:fig3}
\end{figure*}

We now calculate non-Bloch winding numbers by deforming the Brillouin zone using parameters of the bulk-state wave functions.
Replacing $e^{ik}$ with $\lambda_{\alpha,i}$ in $h_{\alpha}(k)$ and defining $p_{\alpha,i}$ as the phase of $\lambda_{\alpha,i}$, we have the non-Bloch Hamiltonian
\begin{equation}
\tilde{H}_{\alpha,i}(k)=\tilde{h}^{x}_{\alpha,i}\sigma_{x}+\tilde{h}^{y}_{\alpha,i}\sigma_{y},
\end{equation}
where
\begin{align}
\label{NBhxhy}
&\tilde{h}^{x}_{\alpha,i}=t_{1}^{\alpha}+t_{2}\frac{|\lambda_{\alpha,i}(p_{\alpha,i})|e^{ip_{\alpha,i}}+\frac{1}{|\lambda_{\alpha,i}(p_{\alpha,i})|}e^{-ip_{\alpha,i}}}{2},\\
&\tilde{h}^{y}_{\alpha,i}=\frac{\gamma}{2}i+t_{2}\frac{|\lambda_{\alpha,i}(p_{\alpha,i})|e^{ip_{\alpha,i}}-\frac{1}{|\lambda_{\alpha,i}(p_{\alpha,i})|}e^{-ip_{\alpha,i}}}{2i}.
\end{align}

\begin{figure*}[tbp]
\center{\includegraphics[width=1\linewidth]{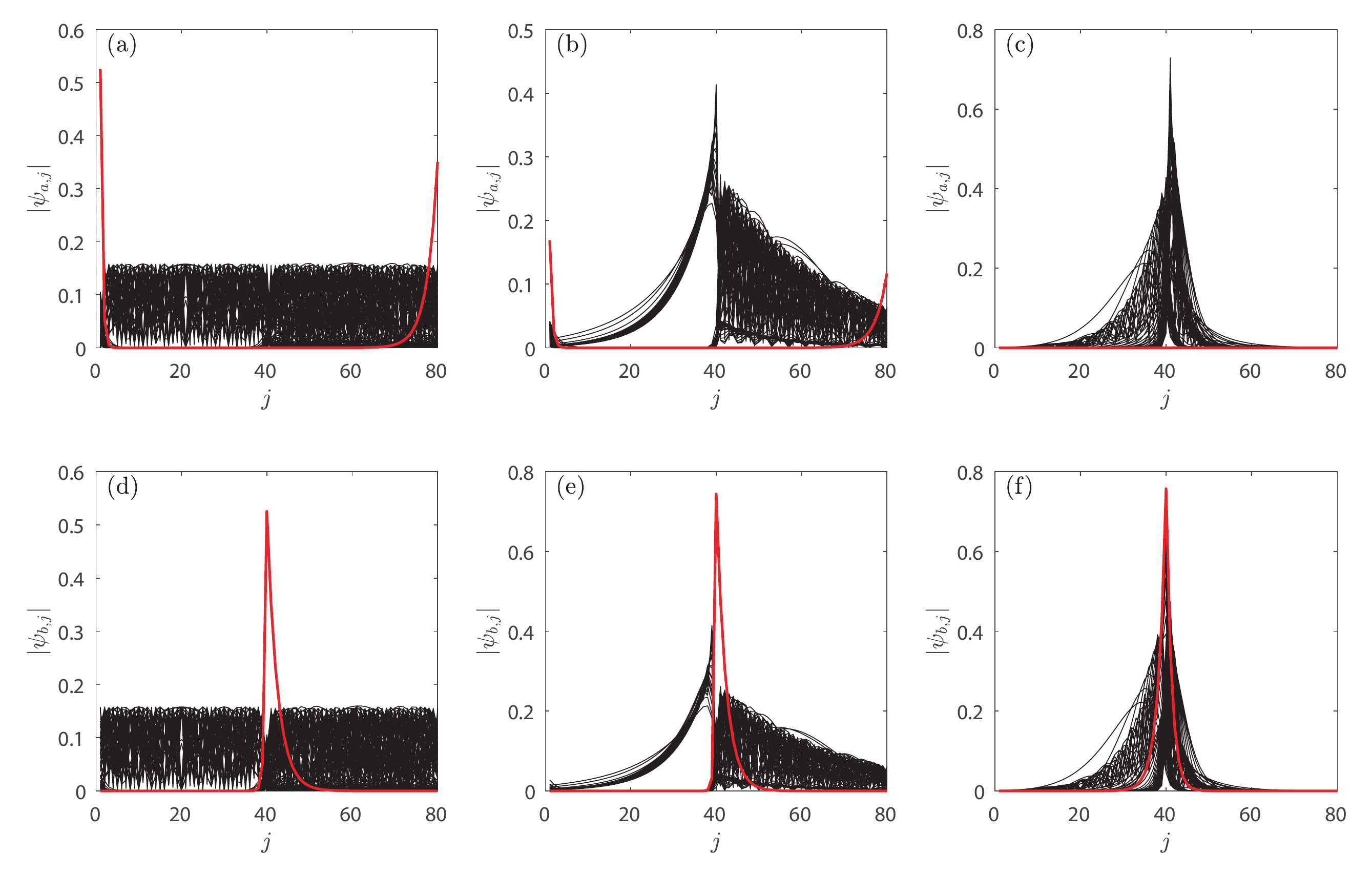}}
\caption{Norm of eigenstate wave functions on sublattice $a$ (upper row) and site $b$ (lower row) with increasing $\gamma$. The black lines correspond to bulk-state wave functions and the red lines correspond to zero-mode wave functions. In (a)(d), we have $\gamma=0$; in (b)(e), $\gamma=0.11t_2$; in (c)(f), $\gamma=1.33t_2$. Other parameters are $t_1^L=-0.1t_2$, $t_1^R=1.5t_2$ and $N_L=N_R=40$.}
\label{fig:fig4}
\end{figure*}

We obtain non-Bloch winding numbers by replacing $h^{x(y)}_\alpha$ with  $\tilde{h}^{x(y)}_\alpha$ in Eq.~(\ref{nualpha})
\begin{equation}
\label{NBin}
\tilde{\nu}_{\alpha,i}=\frac{1}{2\pi}\int dp_{\alpha,i}\frac{-\tilde{h}^{x}_{\alpha,i}\frac{\partial\tilde{h}^{y}_{\alpha,i}}{\partial p_{\alpha,i}}+\tilde{h}^{y}_{\alpha,i}\frac{\partial\tilde{h}^{x}_{\alpha,i}}{\partial p_{\alpha,i}}}{(\tilde{h}^{x}_{\alpha,i})^{2}+(\tilde{h}^{y}_{\alpha,i})^{2}}
\end{equation}
In Appendix~\ref{nu12eq}, we prove $\tilde{\nu}_{\alpha,1}=\tilde{\nu}_{\alpha,2} (\alpha=L,R)$, so that we can denote $\tilde{\nu}_{\alpha,1}= \tilde{\nu}_{\alpha,2}=\tilde{\nu}_{\alpha}$. The two deformed Brillouin zones give the same winding number, consistent with the requirement of bulk-boundary correspondence.

In Figs.~\ref{fig:fig3}(a)(b)(c), we show eigen-energy minima on the $t_1^L $ - $t_1^R$ plane. Zero-energy modes exist in dark blue regions, where finite winding-number difference should exist as dictated by the bulk-edge correspondence. Further, parameters for the right and left bulks are identical when $t_1^L=t_1^R$ (black lines), in which case the system becomes homogeneous. It follows that the Bloch wave-vector $k$ is a good quantum number and the energy spectrum is dictated by the Bloch Hamiltonian Eq.~(\ref{Hksigma}). As we show in Fig.~\ref{fig:fig3}, gapless points along these black lines occur at $t_1^\alpha=\pm t_2\pm\gamma/2$, consistent with those predicted by Eq.~(\ref{Ek}).

In contrast, for the case of $t_1^L=-t_1^R$ (magenta dashed lines), the winding numbers of domain-wall systems become the same as those of open-boundary systems under the same parameters ($t_1^L=-t_1^R$). This is seen by rewriting Eq.~(\ref{beta}) as
\begin{align}
\label{betaeta}
E^{2}&=(t_{1}^{L}-\frac{\gamma}{2}+t_{2}\lambda_{L,i})(t_{1}^{L}+\frac{\gamma}{2}+t_{2}\frac{1}{\lambda_{L,i}}),\\
E^{2}&=(t_{1}^{L}+\frac{\gamma}{2}-t_{2}\lambda_{R,i})(t_{1}^{L}-\frac{\gamma}{2}-t_{2}\frac{1}{\lambda_{R,i}}),
\end{align}
from which we deduce $\lambda_{L,i}=-{1}/\lambda_{R,i}$. Assuming $N$ to be an even number, we reduce Eq.~(\ref{detM}) into
\begin{align}
&(\lambda_{L,1}f_{1}-\lambda_{R,1}g_{1})(\lambda_{L,2}f_{2}-\lambda_{R,2}g_{2})\nonumber\\
&\times[1-(-\frac{\lambda_{L,1}}{\lambda_{L,2}})^{N}][1-(\frac{\lambda_{L,2}}{\lambda_{L,1}})^{N}]=0.
\end{align}
This implies $|\lambda_{L,1}|=|\lambda_{L,2}|$ and $|\lambda_{R,1}|=|\lambda_{R,2}|$ in the limit of $N_L=N_R\rightarrow\infty$, which coincide with conditions of an open-boundary system~\cite{NHtopo3}.
Accordingly, the gapless points on magenta dashed lines are given by $t_{1}^{L}=-t_1^R=\pm\sqrt{t_{2}^{2}+\frac{\gamma^{2}}{4}}$.

The validity of the non-Bloch bulk-boundary correspondence is confirmed in Figs.~\ref{fig:fig3}(d)(e)(f), where we compare non-Bloch and Bloch winding numbers with the numerical energy spectrum. In the Hermitian case $\gamma=0$, non-Bloch and Bloch winding numbers give the same results, as the system has no zero modes when $(|t_1^L|-|t_2|)(|t_1^R|-|t_2|)>0$. As $\gamma$ increases, difference in the non-Bloch winding numbers of the two bulks correctly give the number of topologically protected zero modes, whereas the Bloch winding numbers fail to do so. Further, we note that the non-Bloch winding number of a given bulk typically varies with parameters of the other bulk, an important feature of the domain-wall configuration. For instance, $\tilde{\nu}_R$ changes from $0$ to $0.5$ around $t_1^R\approx 0.16$ in Fig.~\ref{fig:fig3}(f), whereas all parameters of the right bulk stay the same. We note that, due to finite-size effects~\cite{FinNHSSH}, edge-state energies in Fig.~\ref{fig:fig3} not exactly zero and they merge smoothly into the bulk-state spectrum in the vicinity of phase boundaries. However, for our numerical calculation with $N_L=N_R=40$, the system is already sufficiently large such that we are able to confirm the validity of non-Bloch winding numbers despite small finite-size effects.


In Fig.~\ref{fig:fig4}, we plot both bulk-state wave functions and those of zero modes. In the Hermitian case [see Fig.~{\ref{fig:fig4}}(a)(d)], bulk states are extended and zero modes are all localized at the two boundaries. In contrast, in the non-Hermitian case [see Fig.~{\ref{fig:fig4}}(b)(c)(e)(f)], all bulk states are localized. Since localization of bulk-state wave functions can be understood from Eqs.~(\ref{psiL})(\ref{psiR}), when $|\lambda_{L,i}|$ ($|1/\lambda_{R,i}|$) is larger, the corresponding bulk states become more localized. We have numerically confirmed this point for bulk states appearing in Fig.~{\ref{fig:fig4}}.

Further, the localization of zero-mode wave functions are parameter-dependent in the non-Hermitian case. As $\gamma$ increases from zero, the occupation of zero modes near the boundary $j=1,N_L+N_R$ on sublattice site $a$ decreases gradually, whereas for large enough $\gamma$ zero-mode wave functions are completely localized near $j=N_L$.
Conditions for the location of zero modes can be derived by setting $E=0$ in Eq.~(\ref{psialpha1}).
Defining $r_{a(b)}:=\frac{\psi_{a(b),j+1}}{\psi_{a(b),j}}$ for $j\in{J}_\alpha$, we have for the zero modes
\begin{align}
&r_a^\alpha=\frac{(t_{1}^{\alpha}-\frac{\gamma}{2})}{t_{2}},\\
&r_b^\alpha=\frac{t_{2}}{(t_{1}^{\alpha}+\frac{\gamma}{2})}.
\end{align}
Apparently, in the presence of zero modes and when $|r_{a(b)}^L|<1$ or $|r_{a(b)}^R|>1$, the zero modes on sublattice site $a(b)$ are localized around $j=1$ and $j=N_L+N_R$. In contrast, when $|r_{a(b)}^L|>1$ or $|r_{a(b)}^R|<1$, the zero modes on sublattice site $a(b)$ are localized around $j=N_L$.


\section{Non-Bloch topological invariants with $N_R\neq N_L$}
\label{rN}

\begin{figure*}[tbp]
\center{\includegraphics[width=1\linewidth]{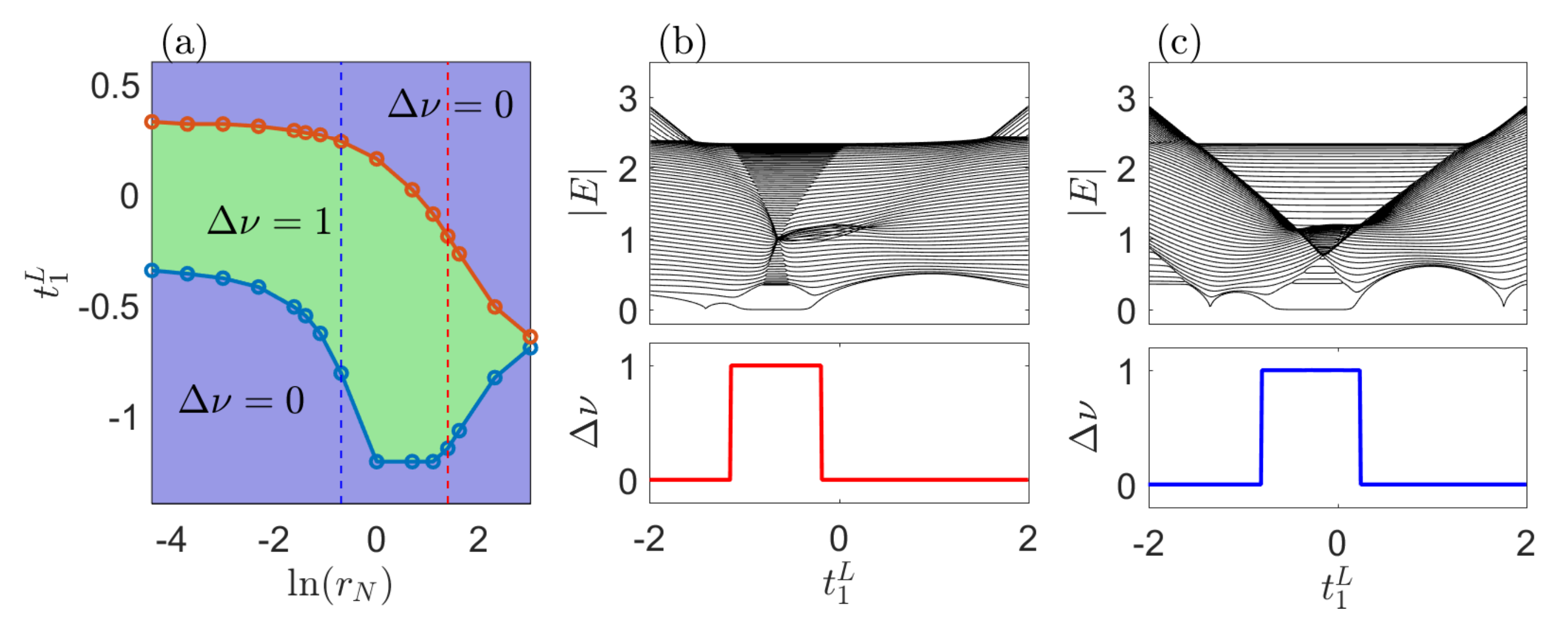}}
\caption{(a)Topological phase diagram in the plane of $t_1^L$--${\rm{ln}}({r}_N)$ with the parameters $t_1^R=1.5t_2$ and $\gamma=1.33t_2$. The
violet region has a vanishing non-Bloch winding-number difference $\Delta\nu=\tilde{\nu}_R-\tilde{\nu}_L=0$, and the green region has a finite difference $\Delta\nu=1$.  The red and blue dashed lines correspond to the parameters in (b) and (c), respectively. (b)(c) The absolute values of the energy spectrum (upper panel) and the non-Bloch winding number difference (lower panel). We take $N_R=4N_L=80$ in (b), and $2N_R=N_L=60$ in (c).}
\label{fig:fig5}
\end{figure*}
In this section, we study non-Bloch winding numbers for $N_R\neq N_L$. We define the length ratio $r_N=N_R/N_L$. For $r_N\neq 1$, we simply replace $\lambda_{R,i}$ with $\lambda_{R,i}^{r_N}$ in Eq.~(\ref{zetaE}) (see Appendix \ref{zetaf}). Then the $\zeta$ function also becomes a function of $r_N$, from which we calculate the $r_N$-dependent winding numbers.
We show the phase diagram of the non-Bloch winding-number difference on the plane of $t_1^R$ - ${\rm{ln}}({r_N})$ in Fig.~\ref{fig:fig5}(a).
In Figs.~\ref{fig:fig5}(b)(c), we confirm that non-Bloch winding numbers still dictate the number of zero modes in the case of $r_N\neq1$.

Notice that in the limit of $r_N\rightarrow\infty$, we can neglect the contribution of $\lambda_{L,i}^{N_L}$ in Eq.~(\ref{detM}). We can then derive the non-Bloch winding number of the right bulk without any information of the left bulk. Eq.~(\ref{zetaE}) is then reduced to $|\lambda_{R,1}|=1$ or $|\lambda_{R,2}|=1$. Therefore, the non-Bloch winding number of the right bulk is the same as the Bloch one. Similarly, when $r_N\rightarrow 0$,
the non-Bloch winding number of the left bulk is the same as the Bloch one as $|\lambda_{L,1}|=1$ or $|\lambda_{L,2}|=1$.

\section{Summary}

We have systematically studied non-Bloch winding numbers and zero modes for a non-Hermitian SSH model in a domain-wall configuration on a ring. Similar to non-Hermitian SSH model with open boundary conditions, the calculation of the non-Bloch winding numbers require information of bulk-state wave functions. However, in contrast to systems with open boundary conditions, here both bulk-state wave function and the non-Bloch winding numbers of either bulk are dependent on parameters of the other bulk. Our results demonstrate the importance of boundary conditions for systems with non-Hermitian skin effects, which would be helpful for experimental studies of non-Bloch bulk-boundary correspondence. They also imply rich phenomena for non-Hermitian domain-wall systems in higher dimensions.

\acknowledgements
We are grateful to Zhong Wang for helpful comments. This work is supported by the The National Key Research and Development Program of China (Grant Nos. 2016YFA0301700,2017YFA0304100).

\appendix
\section{Boundary Condition}
\label{BC}
We need to consider the boundary condition in order to determine $\phi_{a,b}^{(i)}$ ,$\varphi_{a,b}^{(i)}$ (i=1,2) explicitly.
The domain-wall boundary condition is expressed as
\begin{align}\label{Boundary1}
E\psi_{a,1}&=(t_{1}^{L}+\frac{\gamma}{2})\psi_{b,1}+t_{2}\psi_{b,N_{L}+N_{R}},\\
\label{Boundary2}
E\psi_{b,N_{L}}&=(t_{1}^{L}-\frac{\gamma}{2})\psi_{a,N_{L}}+t_{2}\psi_{a,N_{L}+1},\\
\label{Boundary3}
E\psi_{a,N_{L}+1}&=(t_{1}^{R}+\frac{\gamma}{2})\psi_{b,N_{L}+1}+t_{2}\psi_{b,N_{L}},\\
\label{Boundary4}
E\psi_{b,N_{L}+N_{R}}&=(t_{1}^{R}-\frac{\gamma}{2})\psi_{a,N_{L}+N_{R}}+t_{2}\psi_{a,1}.
\end{align}
Substituting Eq.~(\ref{psiL}) and Eq.~(\ref{psiR}) Eqs.~(\ref{Boundary1}-\ref{Boundary4}), we have
\begin{equation}
\label{Mpsi}
M\Psi=0,
\end{equation}
with
\begin{align}
\label{mat}
&M=\\
&\left(\begin{array}{cccc}
-t_{2} & -t_{2} & t_{2}\lambda_{R,1}^{N_{R}} & t_{2}\lambda_{R,2}^{N_{R}}\\
-t_{2}\lambda_{L,1}^{N_{L}+1}f_{1} & -t_{2}\lambda_{L,2}^{N_{L}+1}f_{2} & t_{2}g_{1}\lambda_{R,1} & t_{2}g_{2}\lambda_{R,2}\\
t_{2}\lambda_{L,1}^{N_{L}} & t_{2}\lambda_{L,2}^{N_{L}} & -t_{2} & -t_{2}\\
t_{2}f_{1}\lambda_{L,1} & t_{2}f_{2}\lambda_{L,2} & -t_{2}\lambda_{R,1}^{N_{R}+1}g_{1} & -t_{2}\lambda_{R,2}^{N_{R}+1}g_{2}
\end{array}\right),
\end{align}
and $\Psi=\left[\begin{array}{cccc}
\phi_{b}^{(1)} & \phi_{b}^{(2)} & \varphi_{b}^{(1)} & \varphi_{b}^{(2)}\end{array}\right]^{T}$.
$M$ is the coefficient matrix from which we derive Eq.~(\ref{beta}).

\section{Derivation of $\zeta$}
\label{zetaf}
Starting from Eq.~(\ref{detM}), we first derive Eq.~(\ref{zetaE}) with $N=N_R=N_L$ and $N\rightarrow\infty$.
Without loss of generality, we assume $|\lambda_{L,1}|\geq|\lambda_{L,2}|$
and $|\lambda_{R,1}|\geq|\lambda_{R,2}|$. We then define
\begin{align}
\{ \eta_1&,\eta_2,\eta_3,\eta_4\}
\nonumber\\
&=\{|\lambda_{L,1}\lambda_{R,1}|,|\lambda_{L,1}\lambda_{R,2}|,|\lambda_{L,2}\lambda_{R,1}|,|\lambda_{L,2}\lambda_{R,2}|\},
\end{align}
such that $\eta_1$ is the largest and $\eta_4$ is the smallest. Depending on the detailed ordering of $\eta_m$ ($m=1,2,3,4$) and $1$, Eq.~(\ref{detM}) can be simplified in different ways. In the following, we discuss these situations case-by-case.

In the case of $\eta_1\geqslant1$, $\eta_2\geqslant1$, and $\eta_3\geqslant1$, we transform Eq.~(\ref{detM}) into
\begin{align}\label{Reduction1}
&(\lambda_{L,1}f_{1}-\lambda_{R,1}g_{1})(\lambda_{L,2}f_{2}-\lambda_{R,2}g_{2})(\lambda_{L,2}\lambda_{R,2})^{N}\nonumber\\
&=-(\lambda_{L,1}f_{1}-\lambda_{R,2}g_{2})(\lambda_{L,2}f_{2}-\lambda_{R,1}g_{1})[1-(\lambda_{L,2}\lambda_{R,2})^{N}],
\end{align}
which necessarily leads to $|\lambda_{L,2}\lambda_{R,2}|=1$. This is because if $|\lambda_{L,2}\lambda_{R,2}|>1$ or $|\lambda_{L,2}\lambda_{R,2}|<1$, Eq.~(\ref{Reduction1}) becomes
\begin{align}
&(\lambda_{L,1}f_{1}-\lambda_{R,1}g_{1})(\lambda_{L,2}f_{2}-\lambda_{R,2}g_{2})\nonumber\\
&=(\lambda_{L,1}f_{1}-\lambda_{R,2}g_{2})(\lambda_{L,2}f_{2}-\lambda_{R,1}g_{1}),
\end{align}
or
\begin{align}
(\lambda_{L,1}f_{1}-\lambda_{R,2}g_{2})(\lambda_{L,2}f_{2}-\lambda_{R,1}g_{1})=0,
\end{align}
where the absence of $N$ makes these equations unable to describe all the bulk states~\cite{NHtopo3}.

In the case of $\eta_1\geqslant1$, $\eta_2\geqslant1$, $\eta_3\leqslant1$ and $\eta_4 \leqslant1$, Eq.~(\ref{detM}) reduces to
\begin{align}
&(\lambda_{L,1}f_{1}-\lambda_{R,1}g_{1})(\lambda_{L,2}f_{2}-\lambda_{R,2}g_{2})(\lambda_{R,1})^{N}\nonumber\\
&=(\lambda_{L,1}f_{1}-\lambda_{R,2}g_{2})(\lambda_{L,2}f_{2}-\lambda_{R,1}g_{1})(\lambda_{R,2})^{N}.
\end{align}
It follows that $|\lambda_{R,1}|=|\lambda_{R,2}|$.

In the case of $\eta_1 \geqslant1$, $\eta_3 \geqslant1$, $\eta_2 \leqslant1$ and $\eta_4 \leqslant1$, Eq.~(\ref{detM}) reduces to
\begin{align}
&(\lambda_{L,1}f_{1}-\lambda_{R,1}g_{1})(\lambda_{L,2}f_{2}-\lambda_{R,2}g_{2})(\lambda_{L,1})^{N}\nonumber\\
&=(\lambda_{L,1}f_{1}-\lambda_{R,2}g_{2})(\lambda_{L,2}f_{2}-\lambda_{R,1}g_{1})(\lambda_{L,2})^{N},
\end{align}
which gives $|\lambda_{L,1}|=|\lambda_{L,2}|$.

Finally, in the case of $\eta_2 \leqslant1$, $\eta_3 \leqslant1$ and $\eta_4 \leqslant1$, Eq.~(\ref{detM}) becomes
\begin{align}
&(\lambda_{L,1}f_{1}-\lambda_{R,1}g_{1})(\lambda_{L,2}f_{2}-\lambda_{R,2}g_{2})\nonumber\\
&=(\lambda_{L,1}f_{1}-\lambda_{R,2}g_{2})(\lambda_{L,2}f_{2}-\lambda_{R,1}g_{1})[1-(\lambda_{L,1}\lambda_{R,1})^{N}],
\end{align}
which leads to $|\lambda_{L,1}\lambda_{R,1}|=1$.

Summarizing these four different cases, we define
\begin{align}
&\zeta(\lambda_{L,1},\lambda_{L,2},\lambda_{R,1},\lambda_{R,2})\nonumber\\
&=\begin{cases}
\begin{array}{c}
|\lambda_{L,2}\lambda_{R,2}|-1\\
|\lambda_{R,1}|-|\lambda_{R,2}|\\
|\lambda_{L,1}|-|\lambda_{L,2}|\\
|\lambda_{L,1}\lambda_{R,1}|-1
\end{array} & \begin{array}{c}
\eta_{1}\geqslant1\land\eta_{2}\geqslant1\land\eta_{3}\geqslant1\\
\eta_{1}\geqslant\eta_{2}\geqslant1\geqslant\eta_{3}\geqslant\eta_{4}\\
\eta_{1}\geqslant\eta_{3}\geqslant1\geqslant\eta_{2}\geqslant\eta_{4}\\
\eta_{2}\leqslant1\land\eta_{3}\leqslant1\land\eta_{4}\leqslant1
\end{array}\end{cases},
\end{align}
where the roots of $\zeta(\lambda_{\alpha,i})=0$ give the bulk-state energy spectrum.

As for the case with $N_L\neq N_R$, we rewrite Eq.~(\ref{detM}) in the following form
\begin{align}
&\frac{[1-(\lambda_{L,1}\lambda_{R,1}^{r_{N}})^{N}][1-(\lambda_{L,2}\lambda_{R,2}^{r_{N}})^{N}]}{(\lambda_{L,1}f_{1}-\lambda_{R,1}g_{1})(\lambda_{L,2}f_{2}-\lambda_{R,2}g_{2})}\nonumber\\
&=\frac{[1-(\lambda_{L,1}\lambda_{R,2}^{r_{N}})^{N}][1-(\lambda_{L,2}\lambda_{R,1}^{r_{N}})^{N}]}{(\lambda_{L,1}f_{1}-\lambda_{R,2}g_{2})(\lambda_{L,2}f_{2}-\lambda_{R,1}g_{1})}.
\end{align}
It follows that Eq.~(\ref{zetaE}) can be transformed into
 \begin{align}
\zeta(\lambda_{L,1},\lambda_{L,2},\lambda_{R,1}^{r_N},\lambda_{R,2}^{r_N})=0,
\end{align}
where $r_N =N_R/N_L$.

\section{Proof of $\tilde{\nu}_{\alpha,1}=\tilde{\nu}_{\alpha,2}$}
\label{nu12eq}

\begin{figure}[tbp]
\center{\includegraphics[width=1\linewidth]{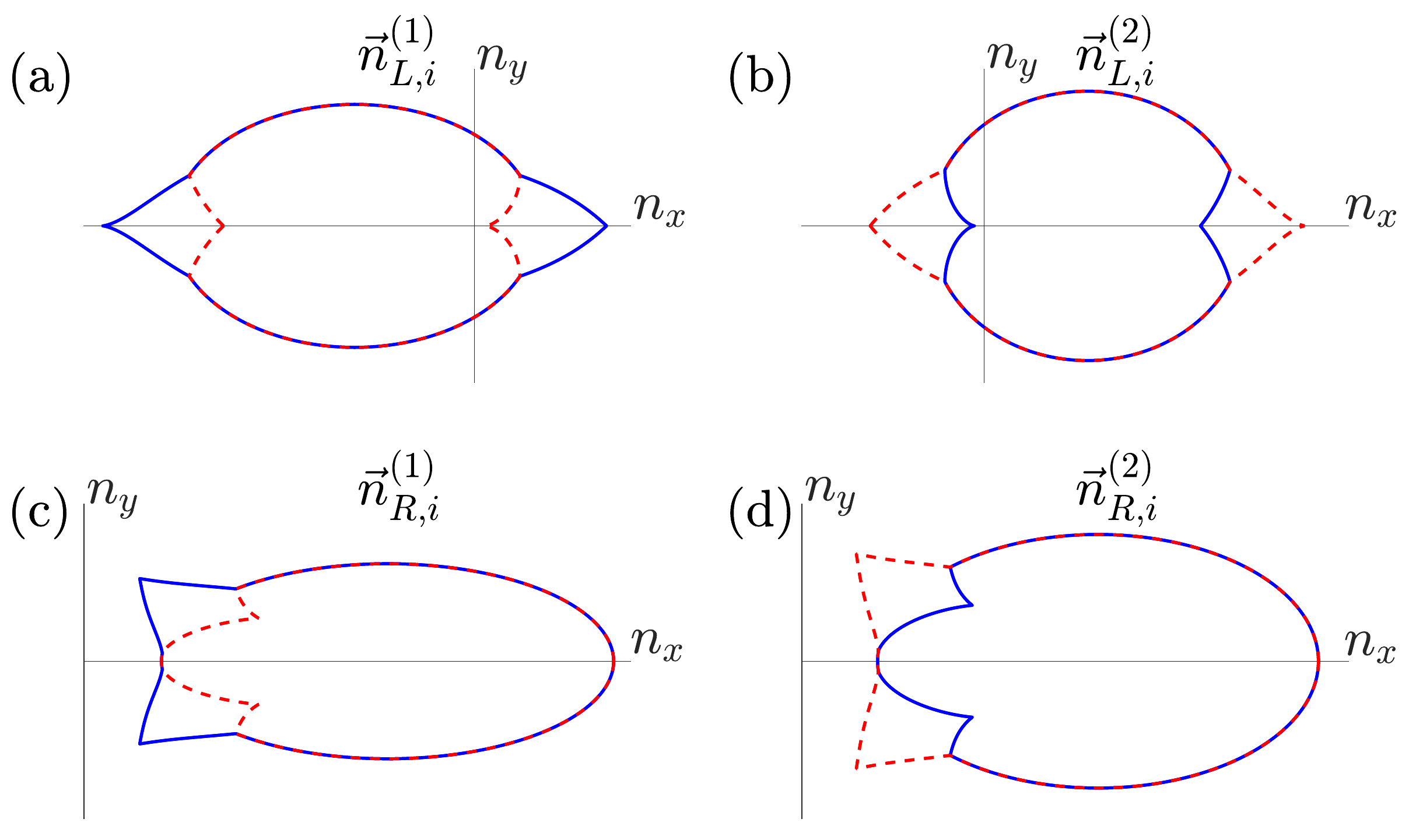}}
\caption{Illustration of $\vec{n}_{\alpha,i}^{(1)}$ and $\vec{n}_{\alpha,i}^{(2)}$, where the blue ($i=1$) and red ($i=2$) lines correspond to vectors associated with distinct non-Bloch Brillouin zones. The $n_x$ ($n_y$) axis represents $x$ ($y$) coordinates of vectors $\vec{n}_{\alpha,i}^{(1)}$ and $\vec{n}_{\alpha,i}^{(2)}$. Here, we take $t_1^R=1.5t_2$, $t_1^L=-0.1t_2$, $\gamma=1.33t_2$, and $r_N=1$.}
\label{fig:figA21}
\end{figure}

In this appendix, we prove that the two different non-Bloch Brillouin zones constructed from the solutions of Eq.~(\ref{beta}) of the same bulk yield the same non-Bloch winding numbers. For this purpose, we start by clarifying the geometrical meaning of non-Bloch winding numbers.

Starting from Eq.~(\ref{NBin}), we adopt the same procedure as in Ref.~\cite{GeoSSH} and split the winding-number integral into two parts
\begin{equation}
\label{In12}
\tilde{\nu}_{\alpha,i}=\frac{\tilde{\nu}_{\alpha,i}^{(1)}+\tilde{\nu}_{\alpha,i}^{(2)}}{2}=\frac{\int d\theta_{\alpha,i}^{(1)}+\int d\theta_{\alpha,i}^{(2)}}{2}.
\end{equation}
Here, the index $i=(1,2)$ denotes the two distinct non-Bloch Brillouin zones, $\alpha$ denotes the left bulk or right bulk, and $\theta_{\alpha,i}^{(1)}$ and $\theta_{\alpha,i}^{(2)}$ are defined as the polar angles of the vectors
\begin{align}
\label{nm1}
&\vec{n}_{\alpha,i}^{(1)}=({\rm Re}\tilde{h}_{\alpha,i}^x-{\rm Im}\tilde{h}_{\alpha,i}^y,{\rm Re}\tilde{h}_{\alpha,i}^y+{\rm Im}\tilde{h}_{\alpha,i}^x),\\
\label{nm2}
&\vec{n}_{\alpha,i}^{(2)}=({\rm Re}\tilde{h}_{\alpha,i}^x+{\rm Im}\tilde{h}_{\alpha,i}^y,{\rm Re}\tilde{h}_{\alpha,i}^y-{\rm Im}\tilde{h}_{\alpha,i}^x).
 \end{align}

Substituting the real and imaginary parts of Eq.~(\ref{NBhxhy}) into Eq.~(\ref{nm1}) and Eq.~(\ref{nm2}), we have
\begin{align}
\label{nm1h}
&\vec{n}_{\alpha,i}^{(1)}=(t_{1}^{\alpha}-\frac{\gamma}{2}+t_{2}|\lambda_{\alpha,i}|\cos p_{\alpha,i},t_{2}|\lambda_{\alpha,i}|\sin p_{\alpha,i}),\\
\label{nm2h}
&\vec{n}_{\alpha,i}^{(2)}=(t_{1}^{\alpha}+\frac{\gamma}{2}+t_{2}\frac{1}{|\lambda_{\alpha,i}|}\cos p_{\alpha,i},t_{2}\frac{1}{|\lambda_{\alpha,i}|}\sin p_{\alpha,i}).
 \end{align}

In Fig.~\ref{fig:figA21}, we show an example of $\vec{n}_{\alpha,i}^{(1)}$ and $\vec{n}_{\alpha,i}^{(2)}$. Geometrically, the winding number $\tilde{\nu}^{(1)}_{\alpha,i}$ ($\tilde{\nu}^{(2)}_{\alpha,i}$) of a given bulk is the number of times $\vec{n}_{\alpha,i}^{(1)}$ ($\vec{n}_{\alpha,i}^{(2)}$) winds around the origin as the corresponding non-Bloch Brillouin zone is traversed. For instance, in Fig.~\ref{fig:figA21}, we have $\tilde{\nu}^{(1)}_{L,i}=\tilde{\nu}^{(2)}_{L,i}=1$ and $\tilde{\nu}^{(1)}_{R,i}=\tilde{\nu}^{(2)}_{R,i}=0$.

To demonstrate the equivalence of different non-Bloch Brillouin zones of a given bulk in defining non-Bloch winding numbers, we further
define $c_\alpha=\frac{t_{1}^{\alpha}-\frac{\gamma}{2}}{t_{1}^{\alpha}+\frac{\gamma}{2}}$. Using Eq.~(\ref{beta}), we have
 \begin{align}
\lambda_{\alpha,1}\lambda_{\alpha,2}=|\lambda_{\alpha,1}|e^{ip_{\alpha,1}}|\lambda_{\alpha,2}|e^{ip_{\alpha,2}}=c_{\alpha},
 \end{align}
which implies $|\lambda_{\alpha,1}||\lambda_{\alpha,2}|=|c_{\alpha}|$ and $p_{\alpha,1}+p_{\alpha,2}=\ln[\text{sign}(c_{\alpha})]$.
We then have
\begin{align}
&\vec{n}_{\alpha,1}^{(1)}=(t_{1}^{\alpha}-\frac{\gamma}{2}+t_{2}|\lambda_{\alpha,1}|\cos p_{\alpha,1},t_{2}|\lambda_{\alpha,1}|\sin p_{\alpha,1})\nonumber\\
&=c_{\alpha}(t_{1}^{\alpha}+\frac{\gamma}{2}+t_{2}\frac{1}{|\lambda_{\alpha,2}|}\cos p_{\alpha,2},t_{2}{{\rm sign}(c_{\alpha})\frac{1}{|\lambda_{\alpha,2}|}\sin p_{\alpha,2}),}
\end{align}
 and
\begin{align}
&\vec{n}_{\alpha,2}^{(1)}=(t_{1}^{\alpha}-\frac{\gamma}{2}+t_{2}|\lambda_{\alpha,2}|\cos p_{\alpha,2},t_{2}|\lambda_{\alpha,2}|\sin p_{\alpha,2})\nonumber\\
&=c_{\alpha}(t_{1}^{\alpha}+\frac{\gamma}{2}+t_{2}\frac{1}{|\lambda_{\alpha,1}|}\cos p_{\alpha,1},t_{2}{\rm sign}(c_{\alpha})\frac{1}{|\lambda_{\alpha,1}|}\sin p_{\alpha,1}).
 \end{align}

It is straightforward to show that $\vec{n}_{\alpha,1}^{(1)}=c_\alpha \vec{n}_{\alpha,2}^{(2)}$ ($c_\alpha \vec{\tilde{n}}_{\alpha,2}^{(2)}$) when $c_\alpha>0$ ($c_\alpha<0$), and $\vec{n}_{\alpha,2}^{(1)}=c_\alpha \vec{n}_{\alpha,1}^{(2)}$( $c_\alpha \vec{\tilde{n}}_{\alpha,1}^{(2)}$) when $c_\alpha>0$ ($c_\alpha<0$). Here, $\vec{\tilde{n}}_{\alpha}$ is defined as the inversion-symmetric counterpart of $\vec{n}_{\alpha}$ with respect to the $x$ axis. From the geometric picture, it is easy to establish that the winding number associated with $\vec{n}_{\alpha,2}^{(2)}$ ($\vec{n}_{\alpha,2}^{(1)}$) is the same as that associated with $\vec{n}_{\alpha,1}^{(1)}$ ($\vec{n}_{\alpha,1}^{(2)}$), such that
   \begin{align}
\tilde{\nu}_{\alpha,1}^{(1)}=\tilde{\nu}_{\alpha,2}^{(2)},\\
\tilde{\nu}_{\alpha,1}^{(2)}=\tilde{\nu}_{\alpha,2}^{(1)}.
 \end{align}
Therefore, we have $\tilde{\nu}_{\alpha,1}=\tilde{\nu}_{\alpha,2}$ according to Eq.~(\ref{In12}). Again, a concrete example is shown in Fig.~\ref{fig:figA21}.

\bibliographystyle{apsrev4-1}
\bibliography{ref}

\begin{thebibliography}{33}%
\makeatletter
\providecommand \@ifxundefined [1]{%
 \@ifx{#1\undefined}
}%
\providecommand \@ifnum [1]{%
 \ifnum #1\expandafter \@firstoftwo
 \else \expandafter \@secondoftwo
 \fi
}%
\providecommand \@ifx [1]{%
 \ifx #1\expandafter \@firstoftwo
 \else \expandafter \@secondoftwo
 \fi
}%
\providecommand \natexlab [1]{#1}%
\providecommand \enquote  [1]{``#1''}%
\providecommand \bibnamefont  [1]{#1}%
\providecommand \bibfnamefont [1]{#1}%
\providecommand \citenamefont [1]{#1}%
\providecommand \href@noop [0]{\@secondoftwo}%
\providecommand \href [0]{\begingroup \@sanitize@url \@href}%
\providecommand \@href[1]{\@@startlink{#1}\@@href}%
\providecommand \@@href[1]{\endgroup#1\@@endlink}%
\providecommand \@sanitize@url [0]{\catcode `\\12\catcode `\$12\catcode
  `\&12\catcode `\#12\catcode `\^12\catcode `\_12\catcode `\%12\relax}%
\providecommand \@@startlink[1]{}%
\providecommand \@@endlink[0]{}%
\providecommand \url  [0]{\begingroup\@sanitize@url \@url }%
\providecommand \@url [1]{\endgroup\@href {#1}{\urlprefix }}%
\providecommand \urlprefix  [0]{URL }%
\providecommand \Eprint [0]{\href }%
\providecommand \doibase [0]{http://dx.doi.org/}%
\providecommand \selectlanguage [0]{\@gobble}%
\providecommand \bibinfo  [0]{\@secondoftwo}%
\providecommand \bibfield  [0]{\@secondoftwo}%
\providecommand \translation [1]{[#1]}%
\providecommand \BibitemOpen [0]{}%
\providecommand \bibitemStop [0]{}%
\providecommand \bibitemNoStop [0]{.\EOS\space}%
\providecommand \EOS [0]{\spacefactor3000\relax}%
\providecommand \BibitemShut  [1]{\csname bibitem#1\endcsname}%
\let\auto@bib@innerbib\@empty
\bibitem [{\citenamefont {Hasan}\ and\ \citenamefont {Kane}(2010)}]{topo1}%
  \BibitemOpen
  \bibfield  {author} {\bibinfo {author} {\bibfnamefont {M.~Z.}\ \bibnamefont
  {Hasan}}\ and\ \bibinfo {author} {\bibfnamefont {C.~L.}\ \bibnamefont
  {Kane}},\ }\href {\doibase 10.1103/RevModPhys.82.3045} {\bibfield  {journal}
  {\bibinfo  {journal} {Rev. Mod. Phys.}\ }\textbf {\bibinfo {volume} {82}},\
  \bibinfo {pages} {3045} (\bibinfo {year} {2010})}\BibitemShut {NoStop}%
\bibitem [{\citenamefont {Qi}\ and\ \citenamefont {Zhang}(2011)}]{topo2}%
  \BibitemOpen
  \bibfield  {author} {\bibinfo {author} {\bibfnamefont {X.-L.}\ \bibnamefont
  {Qi}}\ and\ \bibinfo {author} {\bibfnamefont {S.-C.}\ \bibnamefont {Zhang}},\
  }\href {\doibase 10.1103/RevModPhys.83.1057} {\bibfield  {journal} {\bibinfo
  {journal} {Rev. Mod. Phys.}\ }\textbf {\bibinfo {volume} {83}},\ \bibinfo
  {pages} {1057} (\bibinfo {year} {2011})}\BibitemShut {NoStop}%
\bibitem [{\citenamefont {Ando}(2013)}]{topo3}%
  \BibitemOpen
  \bibfield  {author} {\bibinfo {author} {\bibfnamefont {Y.}~\bibnamefont
  {Ando}},\ }\href {\doibase 10.7566/JPSJ.82.102001} {\bibfield  {journal}
  {\bibinfo  {journal} {Journal of the Physical Society of Japan}\ }\textbf
  {\bibinfo {volume} {82}},\ \bibinfo {pages} {102001} (\bibinfo {year}
  {2013})}\BibitemShut {NoStop}%
\bibitem [{\citenamefont {Bernevig}\ and\ \citenamefont
  {Hughes}(2013)}]{topo4}%
  \BibitemOpen
  \bibfield  {author} {\bibinfo {author} {\bibfnamefont {B.}~\bibnamefont
  {Bernevig}}\ and\ \bibinfo {author} {\bibfnamefont {T.}~\bibnamefont
  {Hughes}},\ }\href {https://books.google.com.hk/books?id=wOn7JHSSxrsC} {\emph
  {\bibinfo {title} {Topological Insulators and Topological Superconductors}}}\
  (\bibinfo  {publisher} {Princeton University Press},\ \bibinfo {year}
  {2013})\BibitemShut {NoStop}%
\bibitem [{\citenamefont {Valenzuela}(2015)}]{topo5}%
  \BibitemOpen
  \bibfield  {author} {\bibinfo {author} {\bibfnamefont {S.~O.}\ \bibnamefont
  {Valenzuela}},\ }\href@noop {} {\emph {\bibinfo {title} {Topological
  insulators: Fundamentals and perspectives}}}\ (\bibinfo  {publisher} {John
  Wiley \& Sons},\ \bibinfo {year} {2015})\BibitemShut {NoStop}%
\bibitem [{\citenamefont {Bansil}\ \emph {et~al.}(2016)\citenamefont {Bansil},
  \citenamefont {Lin},\ and\ \citenamefont {Das}}]{topo6}%
  \BibitemOpen
  \bibfield  {author} {\bibinfo {author} {\bibfnamefont {A.}~\bibnamefont
  {Bansil}}, \bibinfo {author} {\bibfnamefont {H.}~\bibnamefont {Lin}}, \ and\
  \bibinfo {author} {\bibfnamefont {T.}~\bibnamefont {Das}},\ }\href {\doibase
  10.1103/RevModPhys.88.021004} {\bibfield  {journal} {\bibinfo  {journal}
  {Rev. Mod. Phys.}\ }\textbf {\bibinfo {volume} {88}},\ \bibinfo {pages}
  {021004} (\bibinfo {year} {2016})}\BibitemShut {NoStop}%
\bibitem [{\citenamefont {Asb{\'o}th}\ \emph {et~al.}(2016)\citenamefont
  {Asb{\'o}th}, \citenamefont {Oroszl{\'a}ny},\ and\ \citenamefont
  {P{\'a}lyi}}]{topo7}%
  \BibitemOpen
  \bibfield  {author} {\bibinfo {author} {\bibfnamefont {J.~K.}\ \bibnamefont
  {Asb{\'o}th}}, \bibinfo {author} {\bibfnamefont {L.}~\bibnamefont
  {Oroszl{\'a}ny}}, \ and\ \bibinfo {author} {\bibfnamefont {A.}~\bibnamefont
  {P{\'a}lyi}},\ }\href@noop {} {\bibfield  {journal} {\bibinfo  {journal}
  {Lecture notes in physics}\ }\textbf {\bibinfo {volume} {919}} (\bibinfo
  {year} {2016})}\BibitemShut {NoStop}%
\bibitem [{\citenamefont {Lee}(2016)}]{NHtopo1}%
  \BibitemOpen
  \bibfield  {author} {\bibinfo {author} {\bibfnamefont {T.~E.}\ \bibnamefont
  {Lee}},\ }\href {\doibase 10.1103/PhysRevLett.116.133903} {\bibfield
  {journal} {\bibinfo  {journal} {Phys. Rev. Lett.}\ }\textbf {\bibinfo
  {volume} {116}},\ \bibinfo {pages} {133903} (\bibinfo {year}
  {2016})}\BibitemShut {NoStop}%
\bibitem [{\citenamefont {Gong}\ \emph {et~al.}(2018)\citenamefont {Gong},
  \citenamefont {Ashida}, \citenamefont {Kawabata}, \citenamefont {Takasan},
  \citenamefont {Higashikawa},\ and\ \citenamefont {Ueda}}]{NHtopo2}%
  \BibitemOpen
  \bibfield  {author} {\bibinfo {author} {\bibfnamefont {Z.}~\bibnamefont
  {Gong}}, \bibinfo {author} {\bibfnamefont {Y.}~\bibnamefont {Ashida}},
  \bibinfo {author} {\bibfnamefont {K.}~\bibnamefont {Kawabata}}, \bibinfo
  {author} {\bibfnamefont {K.}~\bibnamefont {Takasan}}, \bibinfo {author}
  {\bibfnamefont {S.}~\bibnamefont {Higashikawa}}, \ and\ \bibinfo {author}
  {\bibfnamefont {M.}~\bibnamefont {Ueda}},\ }\href {\doibase
  10.1103/PhysRevX.8.031079} {\bibfield  {journal} {\bibinfo  {journal} {Phys.
  Rev. X}\ }\textbf {\bibinfo {volume} {8}},\ \bibinfo {pages} {031079}
  (\bibinfo {year} {2018})}\BibitemShut {NoStop}%
\bibitem [{\citenamefont {Yao}\ and\ \citenamefont {Wang}(2018)}]{NHtopo3}%
  \BibitemOpen
  \bibfield  {author} {\bibinfo {author} {\bibfnamefont {S.}~\bibnamefont
  {Yao}}\ and\ \bibinfo {author} {\bibfnamefont {Z.}~\bibnamefont {Wang}},\
  }\href {\doibase 10.1103/PhysRevLett.121.086803} {\bibfield  {journal}
  {\bibinfo  {journal} {Phys. Rev. Lett.}\ }\textbf {\bibinfo {volume} {121}},\
  \bibinfo {pages} {086803} (\bibinfo {year} {2018})}\BibitemShut {NoStop}%
\bibitem [{\citenamefont {Yao}\ \emph {et~al.}(2018)\citenamefont {Yao},
  \citenamefont {Song},\ and\ \citenamefont {Wang}}]{NHtopo4}%
  \BibitemOpen
  \bibfield  {author} {\bibinfo {author} {\bibfnamefont {S.}~\bibnamefont
  {Yao}}, \bibinfo {author} {\bibfnamefont {F.}~\bibnamefont {Song}}, \ and\
  \bibinfo {author} {\bibfnamefont {Z.}~\bibnamefont {Wang}},\ }\href {\doibase
  10.1103/PhysRevLett.121.136802} {\bibfield  {journal} {\bibinfo  {journal}
  {Phys. Rev. Lett.}\ }\textbf {\bibinfo {volume} {121}},\ \bibinfo {pages}
  {136802} (\bibinfo {year} {2018})}\BibitemShut {NoStop}%
\bibitem [{\citenamefont {Kunst}\ \emph {et~al.}(2018)\citenamefont {Kunst},
  \citenamefont {Edvardsson}, \citenamefont {Budich},\ and\ \citenamefont
  {Bergholtz}}]{NHtopo5}%
  \BibitemOpen
  \bibfield  {author} {\bibinfo {author} {\bibfnamefont {F.~K.}\ \bibnamefont
  {Kunst}}, \bibinfo {author} {\bibfnamefont {E.}~\bibnamefont {Edvardsson}},
  \bibinfo {author} {\bibfnamefont {J.~C.}\ \bibnamefont {Budich}}, \ and\
  \bibinfo {author} {\bibfnamefont {E.~J.}\ \bibnamefont {Bergholtz}},\ }\href
  {\doibase 10.1103/PhysRevLett.121.026808} {\bibfield  {journal} {\bibinfo
  {journal} {Phys. Rev. Lett.}\ }\textbf {\bibinfo {volume} {121}},\ \bibinfo
  {pages} {026808} (\bibinfo {year} {2018})}\BibitemShut {NoStop}%
\bibitem [{\citenamefont {Martinez~Alvarez}\ \emph
  {et~al.}(2018{\natexlab{a}})\citenamefont {Martinez~Alvarez}, \citenamefont
  {Barrios~Vargas},\ and\ \citenamefont {Foa~Torres}}]{NHtopo6}%
  \BibitemOpen
  \bibfield  {author} {\bibinfo {author} {\bibfnamefont {V.~M.}\ \bibnamefont
  {Martinez~Alvarez}}, \bibinfo {author} {\bibfnamefont {J.~E.}\ \bibnamefont
  {Barrios~Vargas}}, \ and\ \bibinfo {author} {\bibfnamefont {L.~E.~F.}\
  \bibnamefont {Foa~Torres}},\ }\href {\doibase 10.1103/PhysRevB.97.121401}
  {\bibfield  {journal} {\bibinfo  {journal} {Phys. Rev. B}\ }\textbf {\bibinfo
  {volume} {97}},\ \bibinfo {pages} {121401(R)} (\bibinfo {year}
  {2018}{\natexlab{a}})}\BibitemShut {NoStop}%
\bibitem [{\citenamefont {Martinez~Alvarez}\ \emph
  {et~al.}(2018{\natexlab{b}})\citenamefont {Martinez~Alvarez}, \citenamefont
  {Barrios~Vargas}, \citenamefont {Berdakin},\ and\ \citenamefont
  {Foa~Torres}}]{NHtopo8}%
  \BibitemOpen
  \bibfield  {author} {\bibinfo {author} {\bibfnamefont {V.~M.}\ \bibnamefont
  {Martinez~Alvarez}}, \bibinfo {author} {\bibfnamefont {J.~E.}\ \bibnamefont
  {Barrios~Vargas}}, \bibinfo {author} {\bibfnamefont {M.}~\bibnamefont
  {Berdakin}}, \ and\ \bibinfo {author} {\bibfnamefont {L.~E.~F.}\ \bibnamefont
  {Foa~Torres}},\ }\href {\doibase 10.1140/epjst/e2018-800091-5} {\bibfield
  {journal} {\bibinfo  {journal} {The European Physical Journal Special
  Topics}\ }\textbf {\bibinfo {volume} {227}},\ \bibinfo {pages} {1295}
  (\bibinfo {year} {2018}{\natexlab{b}})}\BibitemShut {NoStop}%
\bibitem [{\citenamefont {Lee}\ and\ \citenamefont {Thomale}(2018)}]{NHtopo10}%
  \BibitemOpen
  \bibfield  {author} {\bibinfo {author} {\bibfnamefont {C.~H.}\ \bibnamefont
  {Lee}}\ and\ \bibinfo {author} {\bibfnamefont {R.}~\bibnamefont {Thomale}},\
  }\href@noop {} {\bibfield  {journal} {\bibinfo  {journal} {arXiv preprint
  arXiv:1809.02125}\ } (\bibinfo {year} {2018})}\BibitemShut {NoStop}%
\bibitem [{\citenamefont {Kawabata}\ \emph {et~al.}(2018)\citenamefont
  {Kawabata}, \citenamefont {Shiozaki},\ and\ \citenamefont {Ueda}}]{NHtopo11}%
  \BibitemOpen
  \bibfield  {author} {\bibinfo {author} {\bibfnamefont {K.}~\bibnamefont
  {Kawabata}}, \bibinfo {author} {\bibfnamefont {K.}~\bibnamefont {Shiozaki}},
  \ and\ \bibinfo {author} {\bibfnamefont {M.}~\bibnamefont {Ueda}},\ }\href
  {\doibase 10.1103/PhysRevB.98.165148} {\bibfield  {journal} {\bibinfo
  {journal} {Phys. Rev. B}\ }\textbf {\bibinfo {volume} {98}},\ \bibinfo
  {pages} {165148} (\bibinfo {year} {2018})}\BibitemShut {NoStop}%
\bibitem [{\citenamefont {Herviou}\ \emph {et~al.}(2018)\citenamefont
  {Herviou}, \citenamefont {Bardarson},\ and\ \citenamefont
  {Regnault}}]{NHtopo115}%
  \BibitemOpen
  \bibfield  {author} {\bibinfo {author} {\bibfnamefont {L.}~\bibnamefont
  {Herviou}}, \bibinfo {author} {\bibfnamefont {J.~H.}\ \bibnamefont
  {Bardarson}}, \ and\ \bibinfo {author} {\bibfnamefont {N.}~\bibnamefont
  {Regnault}},\ }\href@noop {} {\bibfield  {journal} {\bibinfo  {journal}
  {arXiv preprint arXiv:1901.00010}\ } (\bibinfo {year} {2018})}\BibitemShut
  {NoStop}%
\bibitem [{\citenamefont {Yokomizo}\ and\ \citenamefont
  {Murakami}(2019)}]{NHtopo12}%
  \BibitemOpen
  \bibfield  {author} {\bibinfo {author} {\bibfnamefont {K.}~\bibnamefont
  {Yokomizo}}\ and\ \bibinfo {author} {\bibfnamefont {S.}~\bibnamefont
  {Murakami}},\ }\href@noop {} {\bibfield  {journal} {\bibinfo  {journal}
  {arXiv preprint arXiv:1902.10958}\ } (\bibinfo {year} {2019})}\BibitemShut
  {NoStop}%
\bibitem [{\citenamefont {Jin}\ and\ \citenamefont {Song}(2019)}]{NHtopo13}%
  \BibitemOpen
  \bibfield  {author} {\bibinfo {author} {\bibfnamefont {L.}~\bibnamefont
  {Jin}}\ and\ \bibinfo {author} {\bibfnamefont {Z.}~\bibnamefont {Song}},\
  }\href {\doibase 10.1103/PhysRevB.99.081103} {\bibfield  {journal} {\bibinfo
  {journal} {Phys. Rev. B}\ }\textbf {\bibinfo {volume} {99}},\ \bibinfo
  {pages} {081103(R)} (\bibinfo {year} {2019})}\BibitemShut {NoStop}%
\bibitem [{\citenamefont {Ghatak}\ and\ \citenamefont {Das}(2019)}]{NHtopo14}%
  \BibitemOpen
  \bibfield  {author} {\bibinfo {author} {\bibfnamefont {A.}~\bibnamefont
  {Ghatak}}\ and\ \bibinfo {author} {\bibfnamefont {T.}~\bibnamefont {Das}},\
  }\href@noop {} {\bibfield  {journal} {\bibinfo  {journal} {arXiv preprint
  arXiv:1902.07972}\ } (\bibinfo {year} {2019})}\BibitemShut {NoStop}%
\bibitem [{\citenamefont {Borgnia}\ \emph {et~al.}(2019)\citenamefont
  {Borgnia}, \citenamefont {Kruchkov},\ and\ \citenamefont
  {Slager}}]{NHtopo17}%
  \BibitemOpen
  \bibfield  {author} {\bibinfo {author} {\bibfnamefont {D.~S.}\ \bibnamefont
  {Borgnia}}, \bibinfo {author} {\bibfnamefont {A.~J.}\ \bibnamefont
  {Kruchkov}}, \ and\ \bibinfo {author} {\bibfnamefont {R.-J.}\ \bibnamefont
  {Slager}},\ }\href@noop {} {\bibfield  {journal} {\bibinfo  {journal} {arXiv
  preprint arXiv:1902.07217}\ } (\bibinfo {year} {2019})}\BibitemShut {NoStop}%
\bibitem [{\citenamefont {Zhou}\ and\ \citenamefont {Lee}(2019)}]{NHtopo18}%
  \BibitemOpen
  \bibfield  {author} {\bibinfo {author} {\bibfnamefont {H.}~\bibnamefont
  {Zhou}}\ and\ \bibinfo {author} {\bibfnamefont {J.~Y.}\ \bibnamefont {Lee}},\
  }\href {\doibase 10.1103/PhysRevB.99.235112} {\bibfield  {journal} {\bibinfo
  {journal} {Phys. Rev. B}\ }\textbf {\bibinfo {volume} {99}},\ \bibinfo
  {pages} {235112} (\bibinfo {year} {2019})}\BibitemShut {NoStop}%
\bibitem [{\citenamefont {Zeuner}\ \emph {et~al.}(2015)\citenamefont {Zeuner},
  \citenamefont {Rechtsman}, \citenamefont {Plotnik}, \citenamefont {Lumer},
  \citenamefont {Nolte}, \citenamefont {Rudner}, \citenamefont {Segev},\ and\
  \citenamefont {Szameit}}]{NHsyn6}%
  \BibitemOpen
  \bibfield  {author} {\bibinfo {author} {\bibfnamefont {J.~M.}\ \bibnamefont
  {Zeuner}}, \bibinfo {author} {\bibfnamefont {M.~C.}\ \bibnamefont
  {Rechtsman}}, \bibinfo {author} {\bibfnamefont {Y.}~\bibnamefont {Plotnik}},
  \bibinfo {author} {\bibfnamefont {Y.}~\bibnamefont {Lumer}}, \bibinfo
  {author} {\bibfnamefont {S.}~\bibnamefont {Nolte}}, \bibinfo {author}
  {\bibfnamefont {M.~S.}\ \bibnamefont {Rudner}}, \bibinfo {author}
  {\bibfnamefont {M.}~\bibnamefont {Segev}}, \ and\ \bibinfo {author}
  {\bibfnamefont {A.}~\bibnamefont {Szameit}},\ }\href {\doibase
  10.1103/PhysRevLett.115.040402} {\bibfield  {journal} {\bibinfo  {journal}
  {Phys. Rev. Lett.}\ }\textbf {\bibinfo {volume} {115}},\ \bibinfo {pages}
  {040402} (\bibinfo {year} {2015})}\BibitemShut {NoStop}%
\bibitem [{\citenamefont {Zhan}\ \emph {et~al.}(2017)\citenamefont {Zhan},
  \citenamefont {Xiao}, \citenamefont {Bian}, \citenamefont {Wang},
  \citenamefont {Qiu}, \citenamefont {Sanders}, \citenamefont {Yi},\ and\
  \citenamefont {Xue}}]{NHsyn1}%
  \BibitemOpen
  \bibfield  {author} {\bibinfo {author} {\bibfnamefont {X.}~\bibnamefont
  {Zhan}}, \bibinfo {author} {\bibfnamefont {L.}~\bibnamefont {Xiao}}, \bibinfo
  {author} {\bibfnamefont {Z.}~\bibnamefont {Bian}}, \bibinfo {author}
  {\bibfnamefont {K.}~\bibnamefont {Wang}}, \bibinfo {author} {\bibfnamefont
  {X.}~\bibnamefont {Qiu}}, \bibinfo {author} {\bibfnamefont {B.~C.}\
  \bibnamefont {Sanders}}, \bibinfo {author} {\bibfnamefont {W.}~\bibnamefont
  {Yi}}, \ and\ \bibinfo {author} {\bibfnamefont {P.}~\bibnamefont {Xue}},\
  }\href {\doibase 10.1103/PhysRevLett.119.130501} {\bibfield  {journal}
  {\bibinfo  {journal} {Phys. Rev. Lett.}\ }\textbf {\bibinfo {volume} {119}},\
  \bibinfo {pages} {130501} (\bibinfo {year} {2017})}\BibitemShut {NoStop}%
\bibitem [{\citenamefont {Xiao}\ \emph {et~al.}(2017)\citenamefont {Xiao},
  \citenamefont {Zhan}, \citenamefont {Bian}, \citenamefont {Wang},
  \citenamefont {Zhang}, \citenamefont {Wang}, \citenamefont {Li},
  \citenamefont {Mochizuki}, \citenamefont {Kim}, \citenamefont {Kawakami}
  \emph {et~al.}}]{NHsyn2}%
  \BibitemOpen
  \bibfield  {author} {\bibinfo {author} {\bibfnamefont {L.}~\bibnamefont
  {Xiao}}, \bibinfo {author} {\bibfnamefont {X.}~\bibnamefont {Zhan}}, \bibinfo
  {author} {\bibfnamefont {Z.}~\bibnamefont {Bian}}, \bibinfo {author}
  {\bibfnamefont {K.}~\bibnamefont {Wang}}, \bibinfo {author} {\bibfnamefont
  {X.}~\bibnamefont {Zhang}}, \bibinfo {author} {\bibfnamefont
  {X.}~\bibnamefont {Wang}}, \bibinfo {author} {\bibfnamefont {J.}~\bibnamefont
  {Li}}, \bibinfo {author} {\bibfnamefont {K.}~\bibnamefont {Mochizuki}},
  \bibinfo {author} {\bibfnamefont {D.}~\bibnamefont {Kim}}, \bibinfo {author}
  {\bibfnamefont {N.}~\bibnamefont {Kawakami}},  \emph {et~al.},\ }\href@noop
  {} {\bibfield  {journal} {\bibinfo  {journal} {Nature Physics}\ }\textbf
  {\bibinfo {volume} {13}},\ \bibinfo {pages} {1117} (\bibinfo {year}
  {2017})}\BibitemShut {NoStop}%
\bibitem [{\citenamefont {Weimann}\ \emph {et~al.}(2017)\citenamefont
  {Weimann}, \citenamefont {Kremer}, \citenamefont {Plotnik}, \citenamefont
  {Lumer}, \citenamefont {Nolte}, \citenamefont {Makris}, \citenamefont
  {Segev}, \citenamefont {Rechtsman},\ and\ \citenamefont {Szameit}}]{NHsyn5}%
  \BibitemOpen
  \bibfield  {author} {\bibinfo {author} {\bibfnamefont {S.}~\bibnamefont
  {Weimann}}, \bibinfo {author} {\bibfnamefont {M.}~\bibnamefont {Kremer}},
  \bibinfo {author} {\bibfnamefont {Y.}~\bibnamefont {Plotnik}}, \bibinfo
  {author} {\bibfnamefont {Y.}~\bibnamefont {Lumer}}, \bibinfo {author}
  {\bibfnamefont {S.}~\bibnamefont {Nolte}}, \bibinfo {author} {\bibfnamefont
  {K.}~\bibnamefont {Makris}}, \bibinfo {author} {\bibfnamefont
  {M.}~\bibnamefont {Segev}}, \bibinfo {author} {\bibfnamefont
  {M.}~\bibnamefont {Rechtsman}}, \ and\ \bibinfo {author} {\bibfnamefont
  {A.}~\bibnamefont {Szameit}},\ }\href@noop {} {\bibfield  {journal} {\bibinfo
   {journal} {Nature materials}\ }\textbf {\bibinfo {volume} {16}},\ \bibinfo
  {pages} {433} (\bibinfo {year} {2017})}\BibitemShut {NoStop}%
\bibitem [{\citenamefont {Zhu}\ \emph {et~al.}(2018)\citenamefont {Zhu},
  \citenamefont {Fang}, \citenamefont {Li}, \citenamefont {Sun}, \citenamefont
  {Li}, \citenamefont {Jing},\ and\ \citenamefont {Chen}}]{NHsyn3}%
  \BibitemOpen
  \bibfield  {author} {\bibinfo {author} {\bibfnamefont {W.}~\bibnamefont
  {Zhu}}, \bibinfo {author} {\bibfnamefont {X.}~\bibnamefont {Fang}}, \bibinfo
  {author} {\bibfnamefont {D.}~\bibnamefont {Li}}, \bibinfo {author}
  {\bibfnamefont {Y.}~\bibnamefont {Sun}}, \bibinfo {author} {\bibfnamefont
  {Y.}~\bibnamefont {Li}}, \bibinfo {author} {\bibfnamefont {Y.}~\bibnamefont
  {Jing}}, \ and\ \bibinfo {author} {\bibfnamefont {H.}~\bibnamefont {Chen}},\
  }\href {\doibase 10.1103/PhysRevLett.121.124501} {\bibfield  {journal}
  {\bibinfo  {journal} {Phys. Rev. Lett.}\ }\textbf {\bibinfo {volume} {121}},\
  \bibinfo {pages} {124501} (\bibinfo {year} {2018})}\BibitemShut {NoStop}%
\bibitem [{\citenamefont {El-Ganainy}\ \emph {et~al.}(2018)\citenamefont
  {El-Ganainy}, \citenamefont {Makris}, \citenamefont {Khajavikhan},
  \citenamefont {Musslimani}, \citenamefont {Rotter},\ and\ \citenamefont
  {Christodoulides}}]{NHsyn4}%
  \BibitemOpen
  \bibfield  {author} {\bibinfo {author} {\bibfnamefont {R.}~\bibnamefont
  {El-Ganainy}}, \bibinfo {author} {\bibfnamefont {K.~G.}\ \bibnamefont
  {Makris}}, \bibinfo {author} {\bibfnamefont {M.}~\bibnamefont {Khajavikhan}},
  \bibinfo {author} {\bibfnamefont {Z.~H.}\ \bibnamefont {Musslimani}},
  \bibinfo {author} {\bibfnamefont {S.}~\bibnamefont {Rotter}}, \ and\ \bibinfo
  {author} {\bibfnamefont {D.~N.}\ \bibnamefont {Christodoulides}},\
  }\href@noop {} {\bibfield  {journal} {\bibinfo  {journal} {Nature Physics}\
  }\textbf {\bibinfo {volume} {14}},\ \bibinfo {pages} {11} (\bibinfo {year}
  {2018})}\BibitemShut {NoStop}%
\bibitem [{\citenamefont {Parto}\ \emph {et~al.}(2018)\citenamefont {Parto},
  \citenamefont {Wittek}, \citenamefont {Hodaei}, \citenamefont {Harari},
  \citenamefont {Bandres}, \citenamefont {Ren}, \citenamefont {Rechtsman},
  \citenamefont {Segev}, \citenamefont {Christodoulides},\ and\ \citenamefont
  {Khajavikhan}}]{NHsyn7}%
  \BibitemOpen
  \bibfield  {author} {\bibinfo {author} {\bibfnamefont {M.}~\bibnamefont
  {Parto}}, \bibinfo {author} {\bibfnamefont {S.}~\bibnamefont {Wittek}},
  \bibinfo {author} {\bibfnamefont {H.}~\bibnamefont {Hodaei}}, \bibinfo
  {author} {\bibfnamefont {G.}~\bibnamefont {Harari}}, \bibinfo {author}
  {\bibfnamefont {M.~A.}\ \bibnamefont {Bandres}}, \bibinfo {author}
  {\bibfnamefont {J.}~\bibnamefont {Ren}}, \bibinfo {author} {\bibfnamefont
  {M.~C.}\ \bibnamefont {Rechtsman}}, \bibinfo {author} {\bibfnamefont
  {M.}~\bibnamefont {Segev}}, \bibinfo {author} {\bibfnamefont {D.~N.}\
  \bibnamefont {Christodoulides}}, \ and\ \bibinfo {author} {\bibfnamefont
  {M.}~\bibnamefont {Khajavikhan}},\ }\href {\doibase
  10.1103/PhysRevLett.120.113901} {\bibfield  {journal} {\bibinfo  {journal}
  {Phys. Rev. Lett.}\ }\textbf {\bibinfo {volume} {120}},\ \bibinfo {pages}
  {113901} (\bibinfo {year} {2018})}\BibitemShut {NoStop}%
\bibitem [{\citenamefont {Zhou}\ \emph {et~al.}(2018)\citenamefont {Zhou},
  \citenamefont {Peng}, \citenamefont {Yoon}, \citenamefont {Hsu},
  \citenamefont {Nelson}, \citenamefont {Fu}, \citenamefont {Joannopoulos},
  \citenamefont {Solja{\v c}i{\'c}},\ and\ \citenamefont {Zhen}}]{NHsyn8}%
  \BibitemOpen
  \bibfield  {author} {\bibinfo {author} {\bibfnamefont {H.}~\bibnamefont
  {Zhou}}, \bibinfo {author} {\bibfnamefont {C.}~\bibnamefont {Peng}}, \bibinfo
  {author} {\bibfnamefont {Y.}~\bibnamefont {Yoon}}, \bibinfo {author}
  {\bibfnamefont {C.~W.}\ \bibnamefont {Hsu}}, \bibinfo {author} {\bibfnamefont
  {K.~A.}\ \bibnamefont {Nelson}}, \bibinfo {author} {\bibfnamefont
  {L.}~\bibnamefont {Fu}}, \bibinfo {author} {\bibfnamefont {J.~D.}\
  \bibnamefont {Joannopoulos}}, \bibinfo {author} {\bibfnamefont
  {M.}~\bibnamefont {Solja{\v c}i{\'c}}}, \ and\ \bibinfo {author}
  {\bibfnamefont {B.}~\bibnamefont {Zhen}},\ }\href {\doibase
  10.1126/science.aap9859} {\bibfield  {journal} {\bibinfo  {journal}
  {Science}\ }\textbf {\bibinfo {volume} {359}},\ \bibinfo {pages} {1009}
  (\bibinfo {year} {2018})}\BibitemShut {NoStop}%
\bibitem [{\citenamefont {Yin}\ \emph {et~al.}(2018)\citenamefont {Yin},
  \citenamefont {Jiang}, \citenamefont {Li}, \citenamefont {L\"u},\ and\
  \citenamefont {Chen}}]{GeoSSH}%
  \BibitemOpen
  \bibfield  {author} {\bibinfo {author} {\bibfnamefont {C.}~\bibnamefont
  {Yin}}, \bibinfo {author} {\bibfnamefont {H.}~\bibnamefont {Jiang}}, \bibinfo
  {author} {\bibfnamefont {L.}~\bibnamefont {Li}}, \bibinfo {author}
  {\bibfnamefont {R.}~\bibnamefont {L\"u}}, \ and\ \bibinfo {author}
  {\bibfnamefont {S.}~\bibnamefont {Chen}},\ }\href {\doibase
  10.1103/PhysRevA.97.052115} {\bibfield  {journal} {\bibinfo  {journal} {Phys.
  Rev. A}\ }\textbf {\bibinfo {volume} {97}},\ \bibinfo {pages} {052115}
  (\bibinfo {year} {2018})}\BibitemShut {NoStop}%
\bibitem [{\citenamefont {Lieu}(2018)}]{NHSSH}%
  \BibitemOpen
  \bibfield  {author} {\bibinfo {author} {\bibfnamefont {S.}~\bibnamefont
  {Lieu}},\ }\href {\doibase 10.1103/PhysRevB.97.045106} {\bibfield  {journal}
  {\bibinfo  {journal} {Phys. Rev. B}\ }\textbf {\bibinfo {volume} {97}},\
  \bibinfo {pages} {045106} (\bibinfo {year} {2018})}\BibitemShut {NoStop}%
\bibitem [{\citenamefont {Chen}\ \emph {et~al.}(2019)\citenamefont {Chen},
  \citenamefont {Chen}, \citenamefont {Zhou},\ and\ \citenamefont
  {Xu}}]{FinNHSSH}%
  \BibitemOpen
  \bibfield  {author} {\bibinfo {author} {\bibfnamefont {R.}~\bibnamefont
  {Chen}}, \bibinfo {author} {\bibfnamefont {C.-Z.}\ \bibnamefont {Chen}},
  \bibinfo {author} {\bibfnamefont {B.}~\bibnamefont {Zhou}}, \ and\ \bibinfo
  {author} {\bibfnamefont {D.-H.}\ \bibnamefont {Xu}},\ }\href {\doibase
  10.1103/PhysRevB.99.155431} {\bibfield  {journal} {\bibinfo  {journal} {Phys.
  Rev. B}\ }\textbf {\bibinfo {volume} {99}},\ \bibinfo {pages} {155431}
  (\bibinfo {year} {2019})}\BibitemShut {NoStop}%
\end{thebibliography}%

\end{document}